\documentclass[preprint,12pt]{elsarticle}
\usepackage{lineno,hyperref}
\modulolinenumbers[5]

\journal{Computer Physics Communications}

\usepackage{lipsum}
\makeatletter
\def\ps@pprintTitle{%
 \let\@oddhead\@empty
 \let\@evenhead\@empty
 \def\@oddfoot{}%
 \let\@evenfoot\@oddfoot}
\makeatother

\usepackage[T1]{fontenc}
\usepackage[latin9]{inputenc}
\usepackage{url}
\usepackage{amsmath}
\usepackage{amssymb}
\usepackage{graphicx}
\usepackage{esint}
\usepackage{bbm}
\usepackage{tikz-cd}
\usepackage[export]{adjustbox}
\makeatletter

\usepackage{enumitem}		

\usepackage{color}
\usepackage{parskip}

\newcommand{\IdentOp}{\mathbbm{1}}

\global\long\def\ket#1{\mbox{\ensuremath{\left|#1\right\rangle }}}
\global\long\def\bra#1{\left\langle #1\right|}
\global\long\def\braket#1#2{\left\langle #1\vphantom{#2}\right|\left.#2\vphantom{#1}\right\rangle }
\global\long\def\ketbra#1#2{\left|#2\vphantom{#1}\right\rangle \left\langle #1\vphantom{#2}\right|}

\usepackage{setspace}

\makeatother

\begin{document}

\begin{frontmatter}

\title{Quantum Dynamics in Phase Space using Projected von Neumann Bases}

\author[xxx]{Shai Machnes\corref{cor1}}
\author[xxx]{Elie Ass{\'e}mat}
\author[yyy]{Henrik R. Larsson}
\author[xxx]{David Tannor}
\address[xxx]{Dept. of Chemical Physics, Weizmann Institute of Science, 76100 Rehovot, Israel}
\address[yyy]{Institut f{\"u}r Physikalische Chemie, Christian-Albrechts-Universit{\"a}t zu Kiel, Olshausenstra{\ss}e 40, D-24098 Kiel, Germany}
\cortext[cor1]{Corresponding author, shai.machnes@gmail.com}

\begin{abstract}
We describe the mathematical underpinnings of the biorthogonal von Neumann method for quantum mechanical simulations (PvB).
In particular, we present a detailed discussion of the important issue of non-orthogonal projection onto subspaces of biorthogonal bases, and how this differs from orthogonal projection. We present various representations of the Schr{\"o}dinger equation in the reduced basis and discuss their relative merits. We conclude with illustrative examples and a discussion of the outlook and challenges ahead for the PvB representation.
\end{abstract}

\end{frontmatter}


\newpage
\tableofcontents{}
\newpage

\section{Introduction}

The quantum dynamics of many physical systems involves multiparticle continuua. Examples range from ionization and high harmonic generation in multielectron atoms to molecular processes such as photodissociation and chemical reactions of polyatomics. The large Hilbert space of these systems makes full simulations extremely challenging. It is therefore of great interest to find a compact representation for quantum dynamics in the continuum where accuracy can be easily controlled.

An attractive option is to represent states as phase space objects, so that only those areas of phase space that are actually occupied are needed for the calculation. The most commonly used phase space representation, the Wigner representation, has proven to be expensive computationally, although there has been some recent progress \cite{Cabrera2015}. An alternative is the Husimi representation, although this too is awkward for direct numerical calculations  \cite{Backer2003}.

A third option for a phase space representation is the von Neumann lattice \cite{von-Neumann-grid-1,von-Neumann-grid-3-Gabor}. This representation consists of a discrete lattice of phase space Gaussians, one per each cell of area $h^D$ where $h$ is Planck's constant and $D$ is the number of degrees of freedom. Initial work using this representation found it to be poorly convergent
\cite{von-Neumann-grid-4-Davis-and-Heller}.  About ten years ago, Poirier and coworkers \cite{poirier1a,poirier1b} found a way to converge the method and have applied it recently to several challenging applications \cite{poirier2a,poirier2b}.  Independently, a few years later, our group discovered a different way to converge the von Neumann lattice, based on modifying the Gaussians to be periodic and band-limited \cite{PvB-1st,Asaf-PRL,Norio}. We named the method PvN, \emph{Periodic von Neumann}. The PvN basis is guaranteed to have exactly the same accuracy as a discrete Fourier representation
\cite{Fourier-grid-1,Fourier-grid-2,Fourier-grid-3,FG-Kosloff}, while having the flexibility of a Gaussian basis set.

Since the PvN basis is non-orthogonal, a key aspect of the mathematical formulation involves the concept of biorthogonal bases.
We define the Biorthogonal von Neumann (BvN) basis as the basis biorthogonal to PvN.  Since a crucial part of our method is to exchange the roles of the PvN and BvN bases we refer to the method more generally as Periodic von Neumann with Biorthogonal exchange or PvB.
In this work we further develop the mathematical underpinnings of the PvB method. In particular, we derive new relations for the reduced Hilbert space associated with non-orthogonal bases when projecting down from the full Hilbert space. Furthermore, we analyze the possible forms of the Schr\"odinger equation in the reduced basis. These developments become crucial when applying the methodology to more challenging systems, such as the helium atom in strong fields\cite{PvB-MCTDH-comparison-paper}.

The paper is organized as follows. In Section \ref{sub:Review-of-PvB} we present a brief review of the PvB approach as applied to  the full Hilbert space. We begin by clarifying the underlying Hilbert space spanned by the Fourier grid method. We then define the von Neumann lattice of Gaussians.  By projecting the von Neumann basis onto the Hilbert space spanned by the Fourier grid we generate the Periodic von Neumann (PvN) basis (Section \ref{sub:The-von-Neumann}). Since the PvN basis is non-orthogonal we can define the basis that is biorthogonal (Section \ref{sub:Bi-orthogonal-bases}).  In Section (\ref{sub:PvN,PvB}) we show that the role of the PvN basis and the biorthogonal basis must be exchanged to obtain a sparse representation. We conclude Section \ref{sub:Review-of-PvB} with a discussion of the possible forms of the Schr{\"o}dinger equation in the PvB representation. Having established the formalism on the full Hilbert space, we turn to the representation on a reduced space.  Section \ref{sub:The-reduced-basis} introduces the reduced PvB method, allowing one to represent only those regions of phase space actually occupied by the current state. We then define the biorthogonal bases for the reduced subspace (Section \ref{sub:Bi-orthogonal-bases-for}) and analyze the projection into the reduced subspace (Section \ref{sub:The-Projector}), highlighting the differences between orthogonal and non-orthogonal projections. In Section \ref{sub:The-TDSE-for-the-reduced_state} we derive the various forms of the Schr{\"o}dinger equation in the reduced representations and discuss approximations and performance considerations.  Section \ref{sec:Examples} gives some illustrative examples and Section \ref{sec:Conclusions-and-outlook} gives an outlook and some ideas for future work.

\section{\label{sub:Review-of-PvB}A Review of PvB}

The PvB representation, developed in\cite{PvB-1st,Asaf-PRL,Norio}, projects the von Neumann lattice of Gaussians (Section \ref{sub:The-von-Neumann}) onto the subspace spanned by the Fourier grid
(Section \ref{sub:The-Fourier-Grid}). As the projected Gaussians form a non-orthogonal basis, we define the basis biorthogonal to the projected Gaussians (Section \ref{sub:Bi-orthogonal-bases}). Together these building blocks define the PvB representation (Section \ref{sub:PvN,PvB}). We then present several alternative but equivalent formulations of the Schr{\"o}dinger equation in the PvB representation (Section \ref{sub:Schroedinger-equation-in-PvB}).

\subsection{\label{sub:The-Fourier-Grid}The Fourier grid}

The pseudospectral Fourier method\cite{FG-Kosloff,FG-2,FG-666} (also known as the periodic sinc DVR (Discrete Variable Representation\cite{FG-3})) is the underlying Hilbert space on which we construct the PvB method. A comprehensive exposition can be found in \cite{David-Textbook}.

Functions with support on a finite segment $x\in\left[0, L\right]$ may be assumed, without loss of generality, to be cyclic, and reside in a Hilbert space spanned by
\begin{equation}
\varphi_{n}\left(x\right)=\frac{1}{\sqrt{L}}\exp\left(2\pi i\frac{x}{L}n\right)=\frac{1}{\sqrt{L}}\exp\left(ik_{n}x\right),\,\,\,\forall n\in\mathbb{Z},\,\,k_{n}=\frac{2\pi}{L}n.\label{eq:spectral-basis}
\end{equation}
Limiting bandwidth to $K$ implies $n\in\left[-n_{\textrm{max}}+1,\ldots, n_{\textrm{max}}\right]$ where $n_{\textrm{max}}:=\left\lfloor \frac{KL}{2\pi}\right\rfloor.$ This defines a rectangular area of phase space of area $2KL$ which is spanned by the \emph{spectral basis}, $\left\{ \varphi_{n}\right\} _{n=1}^{N}$, with $N=2n_{\textrm{max}}$. With an inner product defined as $\left\langle f,g\right\rangle :=\int_{0}^{L}f^{*}\left(x\right)g\left(x\right)dx$ this constitutes the \emph{Fourier grid (FG)} Hilbert space, $\mathcal{H}$.

Any $f\in\mathcal{H}$ can be expanded as $f\left(x\right)=\sum_{n=1}^{N}\left\langle \varphi_{n},f\right\rangle \varphi_{n}\left(x\right)$, allowing us to represent $f$ as a column vector of the expansion coefficients, $\vec{f}_{\varphi}$ $:=$ $\left(\left\langle \varphi_{1},f\right\rangle ,\left\langle \varphi_{2},f\right\rangle ,\ldots\left\langle \varphi_{N},f\right\rangle \right)^{T}$. We shall drop the basis designation when implied by context.

Define the \emph{Fourier grid points,} or \emph{sampling points},
as the set of $N$ equidistant points $\left\{ x_{j}=x_{0}+L\frac{j}{N}\right\} _{j=0}^{N-1}$ with $x_0\in\left[0, \frac{L}{N}\right]$. Define the \emph{pseudo-spectral basis} of $\mathcal{H}$ as the set of \emph{pseudo-spectral functions} $\Theta=\left\{ \theta_{m}\left(x\right)\right\} _{m=1}^{N}\in\mathcal{H}$ such that for $f\left(x\right)\in\mathcal{H}$, $f\left(x\right)=\sum_{m=1}^{N}f\left(x_{m}\right)\theta_{m}\left(x\right)$. $f$ may now be represented by its \emph{sampling vector}, $\vec{f}=$ $\vec{f}_{\theta}=$\\
$\left(f\left(x_{1}\right),f\left(x_{2}\right)\ldots,f\left(x_{N}\right)\right)^{T}$.

By expanding the spectral basis functions in the pseudo-spectral basis, one may derive the explicit form of the latter, $\theta_{m}\left(x\right)=\frac{1}{N}\sum_{n=-n_{\textrm{max}}+1}^{n_{\textrm{max}}}\exp\left(ik_{n}\left(x-x_{m}\right)\right)= e^{i\pi\frac{x-x_{m}}{L}}\frac{\sin\left(N\frac{\alpha}{2}\right)}{N\sin\left(\frac{\alpha}{2}\right)}$, with $\alpha:=2\pi\frac{x-x_m}{L}$ \cite[Supplementary Material 1]{Asaf-PRL}.

These are periodic $sinc$ functions, which are localized around $x_{m}$ with $\theta_m\left(x_n\right)=\delta_{mn}$.
The pseudo-spectral basis functions satisfy the normalization condition $\left\langle \theta_{n},\theta_{m}\right\rangle =\frac{N}{L} \delta_{nm}$ \cite{Note-2}.

Thus, functions in $\mathcal{H}$ may be represented in the pseudo-spectral basis by their sampling vector. Unlike the spectral basis representation, this does not require integration to compute the coefficients. The existence of the pseudo-spectral basis, allowing reconstruction of a function from sampled values, is equivalent to the Shannon-Nyquist theorem for band-limited functions.

The projector into the FG Hilbert space is
\begin{equation}\label{eq:P_into_the_FG}
\mathcal{P}:=\frac{L}{N}\sum_{m=1}^{N}\ketbra{\theta_{m}}{\theta_{m}}=\sum_{n=1}^{N}\ketbra{\varphi_{n}}{\varphi_{n}}.
\end{equation}
When $f\left(x\right)\mathcal{\notin H}$, the projector $\mathcal{P}$ minimizes the distance to the projected
state, and consequently maximizes the overlap, i.e.  $\left\langle f,\mathcal{P}f\right\rangle / \left\Vert \mathcal{P}f \right\Vert \ge \left\langle f,g\right\rangle  / \left\Vert g \right\Vert \,\,\,\forall g\in\mathcal{H}$
(see\cite{Porat}). The pseudo-spectral functions are the projection of the sampling functions into $\mathcal{H}$, i.e. $\mathcal{P}\delta\left(x-x_{m}\right)=\theta_m\left(x\right)$. Defining the \emph{collocation} or \emph{sampling pseudo-projection},
\begin{equation}\label{eq:Q_into_the_FG}
\mathcal{Q}f\left(x\right) := \sum_{m=1}^{N}f\left(x_{m}\right)\theta_{m}\left(x\right),
\end{equation}
generally $\mathcal{Q}f\neq\mathcal{P}f$, unless $f\in\mathcal{H}$. For functions that are almost within $\mathcal{H}$, one may opt to accept the easy-to-compute $\mathcal{Q}$ as an approximation of $\mathcal{P}$, which requires costly integration.

Note that $\mathcal{H}$ and the associated pseudo-spectral functions are just one possible choice for the phase space underlying PvB. See \cite{Asaf-Review,Asaf-non-FG,TuckerCarrington} for alternative possibilities.

\subsection{\label{sub:The-von-Neumann}The von Neumann lattice and its projection onto the Fourier grid (PvN)}

Consider a lattice of Gaussians in the $\left(x,p\right)$ plane. Let $\left(\bar{x}_{i},\bar{p}_{i}\right)$
indicate the center of Gaussian no. $i$, and let $\left(\Delta x,\Delta p\right)$ be the spacing between the lattice sites, with $\frac{\Delta x}{\Delta p}=\frac{\sigma_x}{\sigma_p}$. We define
\begin{equation}
g_{\bar{x}_i,\bar{p}_i}(x)=\left(\frac{1}{2\pi\sigma_{x}^{2}}\right)^{1/4}\exp\left(-\left(\frac{x-\bar{x}_i}{2\sigma_{x}}\right)^{2}+\frac{i}{\hbar}\bar{p}_i(x-\bar{x}_i)\right).
\end{equation}
If the lattice spans the infinite plane in the $x$-$p$ phase space it is known as the \emph{von Neumann lattice} \cite{von-Neumann-grid-1,von-Neumann-grid-2,von-Neumann-grid-3-Gabor}. However, in any real calculation the lattice must be truncated to a finite domain $N_{x}\times N_{p}=N$. This leads to a lack of convergence, as discovered independently in the context of quantum mechanics\cite{von-Neumann-grid-4-Davis-and-Heller} and the time-frequency analog in signal processing \cite{Daubechies}. However, if one projects the von Neumann lattice onto the FG basis one builds in periodic boundary conditions and obtains the same convergence as for the FG.

Define a cyclic formulation of the Gaussians,
\begin{equation}
\textrm{mod}_{L}x:=x-L\left\lfloor \frac{x}{L}\right\rfloor,~~~~
g_{\bar{x}_i,\bar{p}_i}^\textrm{mod}(x):= \left(\mathcal{Q}g_{0,0}(\textrm{mod}_{L}\left(x-\bar{x}_i\right)\right)e^{\frac{i}{\hbar}\bar{p}_i\,\textrm{mod}_{L}\left(x-\bar{x}_i\right)}.
\end{equation}
The cyclic projected Gaussians, known as the \emph{Periodic von Neumann} or \emph{PvN basis}, constitute a \emph{periodic Gabor basis}, where all basis functions are related to each other by shifts in $x$ and $p$ \cite{Note-3}. For the conditioning of the overlap matrices (eq. \ref{eq:Overlap-matrices}) it is beneficial to choose the von Neumann lattice points such that $\bar{x}_i$ are a subset of the FG sample points and  $\bar{p}_i$  are a subset of the spectral basis frequencies.

The representation of the PvN basis functions in the $\theta$ basis is given by
\begin{equation}
G_{jk}:=g_{\bar{x}_{k},\bar{p}_{k}}^{\textrm{mod}}(x_{j}).\label{eq:G_jk}
\end{equation}
The FG defines an area of $\left(2K\right)L=2\pi N$ in phase space. Therefore one may intuitively assign a phase space area of $2\pi$ to each of the $N$ Gaussians of the PvN basis, and consider them \textquotedblleft phase space pixels\textquotedblright{}. Defining $P:=\hbar K$, each such pixel covers a phase space area of $2 \pi \hbar =h$.

\subsection{\label{sub:Bi-orthogonal-bases}Biorthogonal bases}

In this section we consider the general theory of non-orthogonal bases and their biorthogonal bases. In the next section we will specialize to the periodic von Neumann basis and its biorthogonal basis. In anticipation of that section, we use the notation $G$ and $B$ here.

Any set of $N$ linearly independent vectors $\mathcal{G}=\left\{ \ket{g_{k}}\right\} _{k=1}^{N}$
in $\mathcal{H}$ may serve as a non orthogonal basis of $\mathcal{H}$. Let $\mathcal{G}$ be represented in the $\Theta$ orthogonal basis of $\mathcal{H}$ by the invertible matrix $G$. Let $\mathcal{B}$ be a similarly defined non-orthogonal basis of $\mathcal{H}$, represented in $\Theta$ by $B$. The bases $\mathcal{G}$ and $\mathcal{B}$ are considered \emph{biorthogonal bases} (a reciprocal relationship) if
\begin{equation}\label{eq:bi-otrho prelim_B_first}
\braket{b_{j}}{g_{k}}=\delta_{jk}\,\,\,\Longleftrightarrow\,\,\,B^{\dagger}G={\IdentOp}_{N}.
\end{equation}
The relation is reciprocal, i.e.
\begin{equation}\label{eq:bi-otrho prelim_G_first}
\braket{g_{j}}{b_{k}}=\delta_{jk}\,\,\,\Longleftrightarrow\,\,\,G^{\dagger}B={\IdentOp}_{N}.
\end{equation}
If, and only if, $\mathcal{G}$ is orthogonal, then so is $\mathcal{B}$ and $\mathcal{G}=\mathcal{B}$. According to eq. \ref{eq:bi-otrho prelim_G_first}, the rows of $G^{\dagger}$ represent the bra states $\left\{ \bra{g_{k}}\right\} $, while the columns of $B$ represent the ket states. The representation in the $B$ basis of any state $\ket{\psi}\in\mathcal{H}$ associated with the sampling vector $\vec{\psi}$ is
\begin{equation}
\vec{\psi}_{B}:=B^{-1}\vec{\psi} = G^{\dagger}\vec{\psi},\label{eq:psi_B}
\end{equation}
with the elements $\psi_{B}$ denoting the overlaps $\braket{g_{k}}\psi$.

The \emph{completeness relation} for $\mathcal{H}$ is
\begin{equation}\label{eq:GB completeness}
BG^{\dagger}=GB^{\dagger}={\IdentOp}_{N} ~~~~~ \quad\textrm{\ensuremath{\Longleftrightarrow}\quad} ~~~~
\mathcal{P}=\sum_{j=1}^N \ket{b_{j}}\bra{g_{j}}=\sum_{j=1}^N \ket{g_{j}}\bra{b_{j}}.
\end{equation}
By construction, the projector in eq. \ref{eq:GB completeness} spans the full FG Hilbert space, and therefore is equal to the projectors in eq. \ref{eq:P_into_the_FG}.

We define the \emph{overlap matrices} of the $G$ and $B$ bases,respectively:
\begin{equation}
S :=  G^{\dagger}G,~~~~~~ S^{-1}  =  B^{\dagger}B.
\label{eq:Overlap-matrices}
\end{equation}
From these relations it follows that
\begin{equation}
\begin{array}{rcccrclcrcl}
G & = & BS, &  & \vec{\psi}_{B} & = & S\vec{\psi}_{G},\\
B & = & GS^{-1}, &  & \vec{\psi}_{G} & = & S^{-1}\vec{\psi}_{B}.
\end{array}\label{eq:Overlap-matrices1}
\end{equation}

Norms of vectors in $B$ and $G$ are computed via $\left\Vert \psi_{G}\right\Vert ^{2}=\vec{\psi^{\dagger}}_{G}S\vec{\psi}_{G}$ and $\left\Vert \psi_{B}\right\Vert ^{2}=\vec{\psi^{\dagger}}_{B}S^{-1}\vec{\psi}_{B}$. Note that the $G$ and $B$ bases cannot be independently normalized.

\subsection{\label{sub:PvN,PvB}PvB - A sparse representation in the basis biorthogonal to PvN}

Many wavefunctions of interest, whether bound states or traveling
wavepackets, are fairly well-localized in phase space. Therefore
a representation whose coefficients are $\beta_{k}=\braket{g_{k}}{\psi}$,
is expected to be sparse, i.e. to have many near-zero elements.

Consider a Fourier grid Hilbert space $\mathcal{H}$. Using eq. \ref{eq:GB completeness}, any state $\ket{\psi}\in\mathcal{H}$ can be represented as
\begin{equation}\label{eq:psi_expanded_in_G}
\ket{\psi}=\left(\sum_{j=1}^{N}\ket{g_{j}}\bra{b_{j}}\right)\ket{\psi}=\sum_{j=1}^{N}\braket{b_{j}}{\psi}\ket{g_{j}}.
\end{equation}
We shall use the term PvN to denote this representation. Alternatively, any state in $\mathcal{H}$ can be represented as
\begin{equation}\label{eq:psi_expanded_in_B}
\ket{\psi}=\left(\sum_{j=1}^{N}\ket{b_{j}}\bra{g_{j}}\right)\ket{\psi}=\sum_{j=1}^{N}\braket{g_{j}}{\psi}\ket{b_{j}}.
\end{equation}
This defines the \emph{biorthogonal von Neumann representation, PvB}.

Note that while the $\bra{g}$ states are highly localized in phase space, the $\ket{b}$ states are highly delocalized. As a result, the PvB representation is sparse, while the PvN representation is not.  The PvB and PvN bases both span $\mathcal{H}$ and therefore these representations contain exactly the same information. From this point onward, we shall assume all states are in $\mathcal{H}$, and default to the pseudo-spectral representation unless otherwise noted. We will also drop the explicit vector notation for states.

\subsection{\label{sub:Schroedinger-equation-in-PvB}The Schr{\"o}dinger equation in PvB}

We now turn to the form of the time independent and time dependent Schr{\"o}dinger equations
(TISE and TDSE, respectively) in the PvB basis. Starting from the TISE, $H\psi=\lambda\psi$, where $\lambda$
is the energy, and using eqs. \ref{eq:bi-otrho prelim_G_first} and \ref{eq:psi_B}, gives
\begin{equation}
\left(G^{\dagger}HB\right)\psi_{B}=\lambda\psi_{B}. \label{eq:TISE_in_B}
\end{equation}
Similarly, the TDSE $\partial_{t}\psi=-\frac{i}{\hbar}H\psi$ takes the form:
\begin{equation}
\partial_{t}\psi_{B}=-\frac{i}{\hbar}\left(G^{\dagger}HB\right)\psi_{B}.\label{eq:TDSE_in_B}
\end{equation}
The term $G^{\dagger}HB=B^{-1}HB$ is a similarity transformation of $H$ and therefore all eigenvalues are real and all evolutions are unitary.
One can make Hermiticity in eq. \ref{eq:TISE_in_B} explicit by multiplying it from the left by $B^{\dagger}B$ and using eq. \ref{eq:GB completeness}, which transforms it into a generalized eigenvalue problem,
\begin{equation}
\left(B^{\dagger}HB\right)\psi_{B}=\left(B^{\dagger}B\right)\lambda\psi_{B}.
\end{equation}

One may rewrite the TDSE in four distinct ways, all strictly equivalent. We denote a Hamiltonian taking a state in basis $X$ and returning a state in basis $Y$ by $H_{YX}$,
\begin{equation}
\begin{array}{ccccccccccc}
G^{\dagger}HB & = & B^{-1}HB & =: & H_{BB} &  & \longrightarrow &  & \partial_{t}\psi_{B} & = & -\frac{i}{\hbar}\,H_{BB}\,\psi_{B\,},\\
B^{\dagger}HB & = & G^{-1}HB & =: & H_{GB} &  & \longrightarrow &  & \partial_{t}\psi_{B} & = & -\frac{i}{\hbar}\,S\,H_{GB}\,\psi_{B}\,,\\
G^{\dagger}HG & = & B^{-1}HG & =: & H_{BG} &  & \longrightarrow &  & \partial_{t}\psi_{B} & = & -\frac{i}{\hbar}\,H_{BG}\,S^{-1}\,\psi_{B}\,,\\
B^{\dagger}HG & = & G^{-1}HG & =: & H_{GG} &  & \longrightarrow &  & \partial_{t}\psi_{B} & = & -\frac{i}{\hbar}\,S\,H_{GG}\,S^{-1}\,\psi_{B}\,.
\end{array}\label{eq:H_GB_BB_BG_GG}
\end{equation}
Although the forms are mathematically equivalent, they require different computational efforts. $H_{BG}=G^{\dagger}HG$
is quick to compute as the integrations on both sides are with the highly-local Gaussians. This is counter-balanced
by the need to compute $S^{-1}$. This issue shall be revisited in Section \ref{sub:The-TDSE-for-the-reduced_state}.

\section{\label{sub:The-reduced-basis}The reduced Hilbert space}

As discussed above, many of the coefficients of $\psi_{B}$ are negligible,
since the phase space-localized state does not overlap with many of the
localized Gaussians. We therefore define a reduced
Hilbert subspace, spanned by a subset of the $\ket{b}$ vectors, whose coefficients are all above some pre-defined threshold (Section \ref{sub:Definition_of_reduced}).
Next we define the biorthogonal bases for this reduced space (Section
\ref{sub:Bi-orthogonal-bases-for}) and the projector into it (Section
\ref{sub:The-Projector}). Finally we mention a different subspace,
spanned by a subset of $\ket{g}$ vectors (Section \ref{sub:check-check}),
which will be useful in later discussions.

\subsection{\label{sub:Definition_of_reduced}Defining the reduced Hilbert
space, $\widetilde{\mathcal{H}}$}

Consider a Hilbert space $\mathcal{H}$, of dimension $N$, spanned by the orthogonal pseudo-spectral basis $\Theta$
and a set of biorthogonal bases $\mathcal{B}$, $\mathcal{G}$, represented in $\Theta$  by the $N\times N$ matrices $B$ and $G$,
with $G$ being the periodic von Neumann Gaussian lattice, as defined in eq. \ref{eq:G_jk}.
A state $\ket{\psi}$ that is localized in phase space will be sparse in the $B$-representation,
i.e. $\psi_{B}$ will have many near-zero elements. For the sake of notational convenience, let us assume
the first $\widetilde{N}$ coefficients in $\psi_{B}$ are significant,
while the remaining $N-\widetilde{N}$ are negligible. In such a case,
we can save computational resources by reducing the vector $\psi_{B}$
of length $N$ to a vector $\psi_{\widetilde{B}}$ of length $\widetilde{N}\ll N$.

Define the \emph{reduced subspace}, $\widetilde{\mathcal{H}}\subseteq\mathcal{H}$
as the Hilbert space spanned by the first $\widetilde{N}$ vectors of
$B$, i.e. all states whose $B$-basis representation is strictly
zero in entries $\left(\widetilde{N}+1\right)\ldots N$.

\begin{equation}
\mathcal{\widetilde{H}}=\textrm{span}\left(\left\{ \ket{b_{j}}\right\} _{j=1}^{\widetilde{N}}\right).\label{eq:red_subspace_def}
\end{equation}

Note that generally, $\widetilde{\mathcal{H}}$ is quite different from
\begin{equation}
\check{\mathcal{H}}:=\textrm{span}\left(\left\{ \ket{g_{j}}\right\} _{j=1}^{\widetilde{N}}\right),\label{eq:check_Hilbert}
\end{equation}
as is evident when $\widetilde{N}=1$. We shall discuss $\check{\mathcal{H}}$
further in \ref{sub:check-check}.

Define \emph{the complementary subspace},$\bar{\mathcal{H}}$, by
\begin{equation}\label{eq:H-straight-sum}
\mathcal{H}=\widetilde{\mathcal{H}}\oplus\bar{\mathcal{H}}\,\,,
\end{equation}

requiring that \emph{every vector in the complementary subspace $\mathcal{\bar{H}}$
is orthogonal to every vector in the reduced subspace, $\widetilde{\mathcal{H}}$}.
As the $B$ basis is non-orthogonal, there is no partitioning of the $B$ functions that allows one to span both subspaces separately. However, the complementary space may be written as
\begin{equation}
\bar{\mathcal{H}}=\textrm{span}\left(\left\{ \ket{g_{k}}\right\} _{k=\widetilde{N}+1}^{N}\right).
\end{equation}

By construction, $\widetilde{\mathcal{H}}$ is rank $\widetilde{N}$, $\bar{\mathcal{H}}$
is rank $N-\widetilde{N}$, and $\widetilde{\mathcal{H}}\perp\bar{\mathcal{H}}$
due to the biorthogonality of their respective base vectors. Therefore
$\textrm{rank}\left(\widetilde{\mathcal{H}}\right)+\textrm{rank}\left(\bar{\mathcal{H}}\right)=\textrm{rank}\left(\mathcal{H}\right)$
and we conclude that indeed $\mathcal{H}=\widetilde{\mathcal{H}}\oplus\bar{\mathcal{H}}$.
Given any state $\ket{\psi}\in\mathcal{H}$, we may therefore decompose
it as
\begin{equation}
\ket{\psi}=\ket{\widetilde{\psi}}+\ket{\bar{\psi}}\textrm{ \,\,\ s.t. }\braket{\widetilde{\psi}}{\bar{\psi}}=0\textrm{\,\,\ and\,\,}\ket{\widetilde{\psi}}\in\widetilde{\mathcal{H}}\textrm{, }\ket{\bar{\psi}}\in\mathcal{\bar{H}},
\end{equation}

where $\ket{\widetilde{\psi}}$ is the \emph{reduced state}
and $\ket{\bar{\psi}}$ \emph{the complementary state}. Projection
may be viewed as subtracting $\ket{\bar{\psi}}$ from $\ket{\psi}$, a view that
will be made explicit in Section \ref{sub:The-Projector}.

\subsection{\label{sub:Bi-orthogonal-bases-for}Biorthogonal bases for $\widetilde{\mathcal{H}}$}

We now turn to the biorthogonal bases for the reduced subspace,
$$\widetilde{\mathcal{H}} = \left\{ \ket{b_{j}}\right\} _{j=1}^{\widetilde{N}}.$$ We define these bases in a manner similar to that of the unreduced bases. Let us denote the first $\widetilde{N}$ columns of the $B$ matrix with the matrix
$\widetilde{B}_{N\times\widetilde{N}}$, i.e. $\widetilde{B}$ is the $\Theta$
representation of the basis defining $\widetilde{\mathcal{H}}$. $\widetilde{G}$ will be defined as a basis which is biorthogonal to
$\widetilde{B}$, and spans the same Hilbert space as $\widetilde{B}$;
i.e. the pair $\widetilde{G}$ and $\widetilde{B}$ satisfy the completeness
relation on the subspace. We stress that $\widetilde{B}$ is the natural basis of  $\widetilde{\mathcal{H}}$, while $\widetilde{G}$ is defined in terms of $\widetilde{B}$. One should note that while $\tilde{B}$ is a truncated version of $B$, the $\tilde{G}$
matrix is not a simple truncation of $G$, but rather that individual $g$ functions are modified by a projection
into the subspace spanned by $\tilde{B}$. This will be discussed in detail later in this section.

The biorthogonality requirement is
\begin{equation}
\widetilde{G}^{\dagger}\widetilde{B}={\IdentOp}_{\widetilde{N}}.\label{eq:red_bi_ortho_req}
\end{equation}

The requirement that $\widetilde{B}$ and $\widetilde{G}$ span the same subspace, is equivalent to the statement that $\widetilde{G}$ vectors are expressible as linear combinations of $\widetilde{B}$ vectors, and vice versa, i.e.

\begin{equation}
\exists C\textrm{ s.t. }\widetilde{B}=\widetilde{G}C,\label{eq:red_completeness_req_1}
\end{equation}

\begin{equation}
\exists D\textrm{ s.t. }\widetilde{G}=\widetilde{B}D.\label{eq:red_completeness_req_2}
\end{equation}
(i.e. $\ket{\widetilde{g}_{1}}=\sum_{k=1}^{\widetilde{N}}C_{k,1}\ket{\widetilde{b}_{k}}$,
etc). The biorthogonality requirement is satisfied by the right pseudo-inverse
of $\widetilde{B}^\dagger$,
\begin{equation}
\widetilde{G}:=\widetilde{B}\left(\widetilde{B}^{\dagger}\widetilde{B}\right)^{-1},\label{eq:G_tilde_def}
\end{equation}

and eq. \ref{eq:red_completeness_req_1} and \ref{eq:red_completeness_req_2}
are satisfied by $D=\widetilde{S}$ where

\begin{equation}
\widetilde{S}:=\left(\widetilde{B}^{\dagger}\widetilde{B}\right)^{-1}=\widetilde{G}^{\dagger}\widetilde{G},\label{eq:tilde_S}
\end{equation}

and $C=\widetilde{S}^{-1}$ where

\begin{equation}
\widetilde{S}^{-1}=\widetilde{B}^{\dagger}\widetilde{B}=\left(\widetilde{G}^{\dagger}\widetilde{G}\right)^{-1}.\label{eq:tilde_S_and_invS}
\end{equation}

Therefore $\widetilde{G}$ and $\widetilde{B}$ are biorthogonal bases, both
spanning $\widetilde{\mathcal{H}}$. As biorthogonality is a reciprocal relation, eq. \ref{eq:G_tilde_def}
is echoed by $\widetilde{B}=\widetilde{G}\left(\widetilde{G}^{\dagger}\widetilde{G}\right)^{-1}$. Using eq. \ref{eq:tilde_S} and \ref{eq:tilde_S_and_invS}, we may relate $\widetilde{G}$ and $\widetilde{B}$ using $\widetilde{S}$: $\widetilde{G}=\widetilde{B}\widetilde{S}$ and $\widetilde{B}=\widetilde{G}{{\widetilde{S}}^{-1}}$.

Note that the biorthogonality requirement alone, eq. \ref{eq:red_bi_ortho_req}, is insufficient to uniquely define $\widetilde{G}$. For example, both the Gaussians, $\ket{g}$, and the vectors $\ket{\widetilde{g}}$ of $\widetilde{G}$ are biorthogonal to the $\ket{b}=\ket{\widetilde{b}}$ vectors spanning the reduced subspace, i.e. $\braket{g_{k}}{b_{j}}=\braket{g_{k}}{\widetilde{b}_{j}}=\delta_{jk}=\braket{\widetilde{g}_{k}}{\widetilde{b}_{j}}\,\,\,\forall j,k\in1 \ldots \widetilde{N}$. But while $\left\{ \ket{\widetilde{b}_{j}}\right\}_{j=1}^{\widetilde{N}}$ and $\left\{ \ket{\widetilde{g}_{j}}\right\}_{j=1}^{\widetilde{N}}$ span the same subspace, $\left\{ \ket{b_{j}}\right\} _{j=1}^{\widetilde{N}}$ and $\left\{ \ket{g_{j}}\right\} _{j=1}^{\widetilde{N}}$ do not (except when $\widetilde{N}=N$). At the extreme, when $\widetilde{N}=1$, it is clear that $\ket{b_1}$ and $\ket{g_1}$ span different (trivial) Hilbert spaces.

Similarly to the definitions of $\widetilde{B}$ and $\widetilde{G}$, we define the basis for the complementary Hilbert space $\bar{\mathcal{H}}$ with $\bar{G}$ of size $N\times\left(N-\widetilde{N}\right)$ as the last $N-\widetilde{N}$
columns of $G$, and $\bar{B}:=\bar{G}\left(\bar{G}^{\dagger}\bar{G}\right)^{-1}$.

\subsection*{The deformation of the $\widetilde{G}$ functions and the $\widetilde{S}$ matrix on the reduced subspace}

\begin{figure}
\noindent \begin{centering}
\includegraphics[scale=0.6]{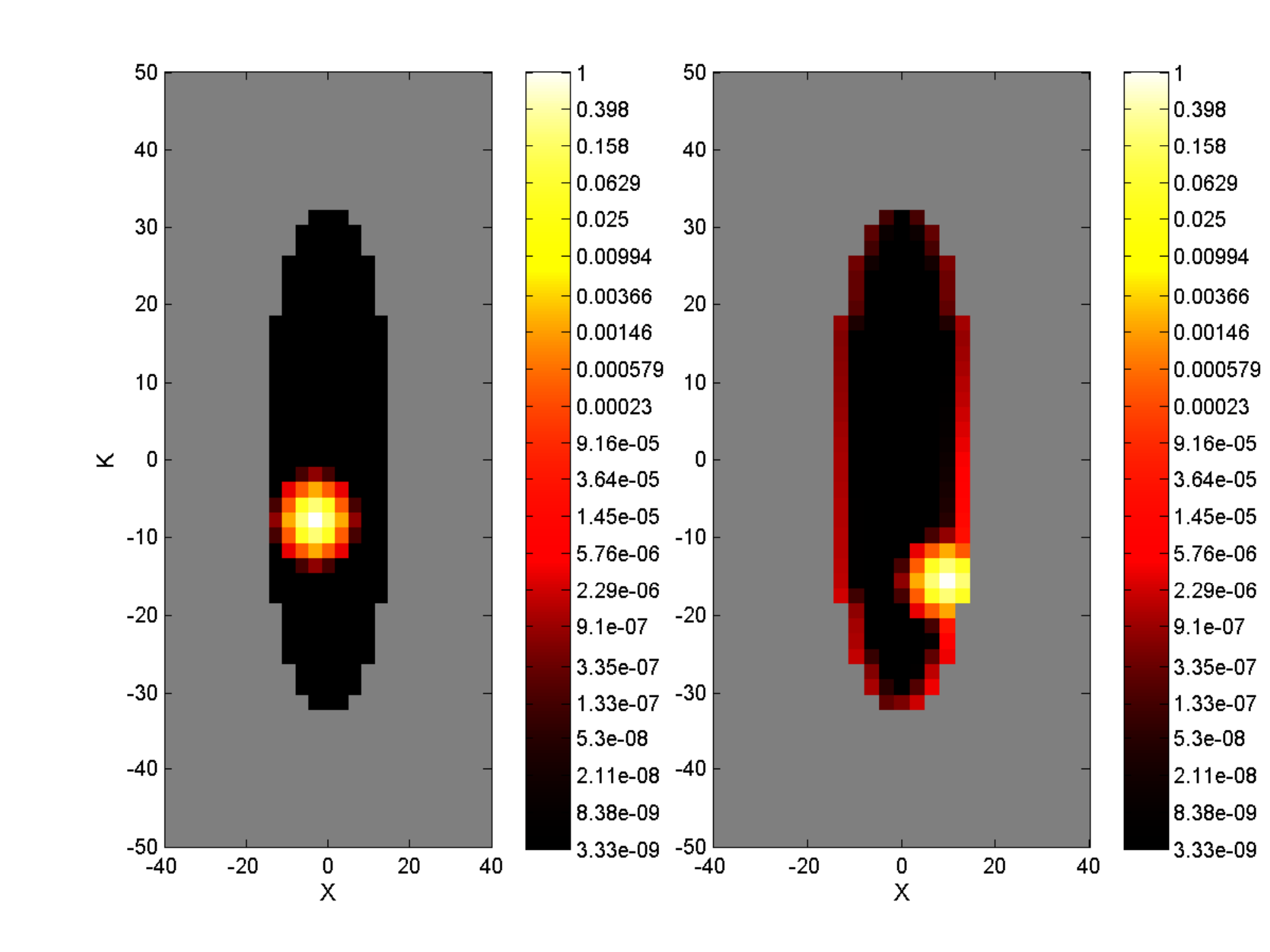}
\par\end{centering}
\noindent \centering{}\protect\caption{Depiction of modified Gaussians associated with the reduced basis. The reduced basis is the non-gray area in both plots. On the left, the modified Gaussian $\widetilde{g}$, which is centered around a non-edge location in the phase space, is almost unchanged. On the right, we see a heavily deformed Gaussian, i.e. the function is delocalized throughout the edge region. This results from the projection of the $\tilde{G}$ basis functions into the subspace spanned by $\tilde{B}$. The deformation is much more significant for Gaussians whose centers are near the edge. See eq. \ref{eq:I_am_so_tired} and the detailed discussion following eq. \ref{eq:tildeP_subtractive_on_gk} for an explanation The states are plotted as heat maps, where the value of each cell of the von Neumann lattice is the absolute value of the overlap of the state plotted (here the modified Gaussians), with the Gaussian centered at that cell of the lattice, $\left|{\left\langle{g_{\bar{x}_{j},\bar{k}_{j}}}\right|\left.\widetilde{g}\right\rangle}\right|$.\label{fig:deformed_Gaussians}}
\end{figure}

The discussion above, and particularly eq. \ref{eq:G_tilde_def}, leads
to the conclusion that the columns of $\widetilde{G}$ are no longer
exactly Gaussians, but have been deformed as the result of the
basis reduction. This is depicted in fig. \ref{fig:deformed_Gaussians}.
Note that Gaussians that are far away from the reduction boundary are not as affected as the
Gaussians closer to the boundary. This is consistent with the intuitive notion that the localized Gaussians should not be affected by changes in areas of phase space into which they do not significantly extend. This intuition is made precise in the following section.

Schematically, the reduced $\widetilde{S}$ matrix is constructed from the $S$ matrix
by the following circuitous path,

\begin{equation}\label{eq:S-flow-diag}
\includegraphics[scale=0.55]{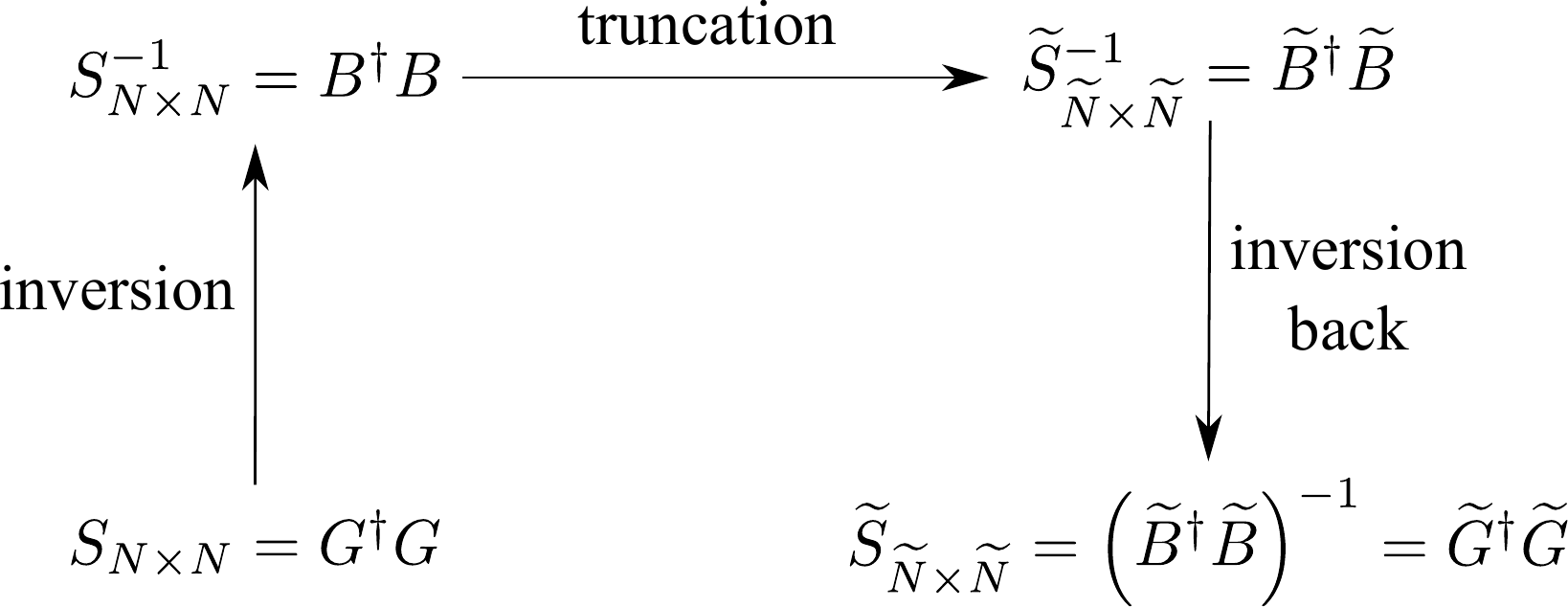}
\end{equation}

One starts with the unmodified $S_{N\times N}$, inverts to get $S^{-1}_{N\times N}$, truncates to get $\widetilde{S}^{-1}_{\widetilde{N}\times\widetilde{N}}$, and then inverts back to get $\widetilde{S}_{\widetilde{N}\times\widetilde{N}}$. The chain of relationships suggest that although $G$ is simpler than $B$, $\widetilde{G}$ is more complicated than $\widetilde{B}$. Specifically, in the multi-dimensional case the columns of the $\widetilde{G}$ matrix do not decompose into products of their one-dimensional counterparts, while the columns of the $\widetilde{B}$ matrix do decompose into $1D$ components. The reduced $\widetilde{S}^{-1}=\widetilde{B}^{\dagger}\widetilde{B}$ matrix is a subset (of rows and columns) of the unreduced $S^{-1}$ matrix, but the reduced $\widetilde{S}=\widetilde{G}^{\dagger}\widetilde{G}$ matrix is not a subset of the full $S$, due to the inversion in eq. \ref{eq:S-flow-diag}. Non-decomposability of the multi-dimensional $\widetilde{S}$ and $\widetilde{G}$ matrices has significant impact on the computational resources required to compute the reduced Hamiltonian. The issue is discussed in depth in Section \ref{sub:The-TDSE-for-the-reduced_state}.

We summarize the identities involving the reduced $\widetilde{S}$ matrix that are counterparts of the identities for the unreduced $S$ matrix, eq. \ref{eq:Overlap-matrices}:
\begin{equation}
\begin{array}{rcccrclcrcl}
\widetilde{S} & := & \widetilde{G}^{\dagger}\widetilde{G}, &  & \widetilde{G} & = & \widetilde{B}\widetilde{S}, &  & \vec{\psi}_{\widetilde{B}} & = & \widetilde{S}\vec{\psi}_{\widetilde{G}},\\
\widetilde{S}^{-1} & = & \widetilde{B}^{\dagger}\widetilde{B}, &  & \widetilde{B} & = & \widetilde{G}\widetilde{S}^{-1}, &  & \vec{\psi}_{\widetilde{G}} & := & \widetilde{S}^{-1}\vec{\psi}_{\widetilde{B}}.
\end{array}\label{eq:Overlap-matrices-red}
\end{equation}

\subsection{\label{sub:The-Projector}Projecting into $\widetilde{\mathcal{H}}$}

We can now expand a state using the reduced basis, echoing eq. \ref{eq:psi_expanded_in_B},
\begin{equation}
\ket{\widetilde{\psi}}:=\sum_{j=1}^{\widetilde{N}}\braket{\widetilde{g}_{j}}{\psi}\ket{\widetilde{b}_{j}}=\left(\sum_{j=1}^{\widetilde{N}}\ketbra{\widetilde{g}_{j}}{\widetilde{b}_{j}}\right)\ket{\psi},
\end{equation}
providing us with a representation-free form of the projector from $\mathcal{H}$ to $\widetilde{\mathcal{H}}$. In Dirac notation,
\begin{equation}\label{eq:P_Dirac}
\mathcal{\widetilde{P}}:=\sum_{j=1}^{\widetilde{N}}\ketbra{\widetilde{g}_{j}}{\widetilde{b}_{j}}=\sum_{j=1}^{\widetilde{N}}\ketbra{\widetilde{b}_{j}}{\widetilde{g}_{j}}\,\,.
\end{equation}
with the second equation following by Hermiticy of the projection. Note similarity to eqs. \ref{eq:psi_expanded_in_G} and \ref{eq:psi_expanded_in_B}. Recall that $\ket{\widetilde{b}_{j}}=\ket{b_{j}}\,\,\,\forall j=1\ldots\widetilde{N}$.

The projector in the $\Theta$ representation (i.e. both input and output are represented on the Fourier grid) is given by

\begin{equation}
\widetilde{P}:=\widetilde{B}\widetilde{G}^{\dagger},\label{eq:P_in_FG}
\end{equation}

where we omit the basis designation $\Theta$. One may directly show that $\widetilde{P}$ is idempotent, i.e. $\widetilde{P}^{2}=\widetilde{P}$, and Hermitian. Moreover, the projector $\widetilde{P}$ transforms a general state to the state closest to it within the subspace \cite{Porat}. The idempotent property of $\widetilde{P}$ requires that the bra matrix in eq. \ref{eq:P_in_FG}, $\widetilde{G}^{\dagger}$, be biorthogonal to the ket matrix, $\widetilde{B}$. Therefore, the projector is non-orthogonal if and only if the bases are non-orthogonal, which is the case if and only if $\widetilde{B}$ and $\widetilde{G}$ are not identical. The projection operator may be rewritten in several representations. We shall denote by $\widetilde{P}_{YX}$ the projection operator from the full Hilbert space $\mathcal{H}$ into the reduced subspace $\widetilde{\mathcal{H}}$, where the Fourier grid representation of the input basis for $\mathcal{H}$ is the matrix $X$, and the Fourier grid representation for $\widetilde{\mathcal{H}}$ is the matrix $Y$. $\widetilde{P}_{YX}$ may be square or rectangular, depending on the choice of output basis. The projection operator transforms as do all operators, using

\begin{equation}
P_{YX}=Y^{-1}PX. \label{eq:P_YX}
\end{equation}

When $Y$ is not square, $Y^{-1}$ should be taken to be the left pseudo-inverse of $Y$.

To better understand the difference between this non-orthogonal projector and a standard orthogonal projector we shall examine two representations: $\widetilde{P}_{\widetilde{B}B}$ where the projection operation will be viewed as an additive operation, and $\widetilde{P}_{\widetilde{G}G}$, where the projection operation will be viewed as a subtractive operation.

\subsection*{$\widetilde{P}_{\widetilde{B}B}$ - All basis vectors contribute to the reduced subspace}

Given a state in the $B$ basis, $\ket{\psi}=\sum_{k=1}^{N} \beta_k \ket{b_k}$, the application to $\ket{\psi}$ of eq. \ref{eq:P_Dirac}  gives
\begin{equation}
\mathcal{\widetilde{P}}\ket{\psi} = \sum_{j=1}^{\widetilde{N}} \sum_{k=1}^{N} \beta_k \ket{\widetilde{g}_{j}}\braket{\widetilde{b}_{j}} {b_k}.
\label{eq:P_tildeB_B_applied}\end{equation}
Note that all $b_k$ vectors contribute to the coefficients in the reduced subspace, as indicated by the $\braket{\widetilde{b}_{j}}{b_k}$ term of eq. \ref{eq:P_tildeB_B_applied}.  The participation of all $b_k$ vectors in the coefficients is reflected in the $\widetilde{P}_{\widetilde{B}B}$ representation of the projector, as depicted in fig. \ref{fig:P_BB}. We derive the explicit form of $\widetilde{P}_{\widetilde{B}B}$ from eq.
\ref{eq:P_YX}, using eq. \ref{eq:red_bi_ortho_req}, giving
\begin{equation}
\widetilde{P}_{\widetilde{B}B}=\widetilde{G}^{\dagger}\widetilde{P}B.
\label{eq:P_BB_1}\end{equation}
Examining the top panel of fig. \ref{fig:P_BB}, the right block of $\widetilde{P}_{\widetilde{B}B}$ depicts the overlap of all $B$ vectors of the unreduced basis, including $\left\{ \ket{b_{j}}\right\} _{j=\widetilde{N}+1}^{N}$, with the $\left\{ \ket{b_{j}}\right\} _{j=1}^{\widetilde{N}}$ subset. This added contribution is unique to non-orthogonal projections. If this were an orthogonal projection, only the basis vectors defining the reduced basis, $\left\{ \ket{b_{j}}\right\} _{j=1}^{\widetilde{N}}$, which make up the identity matrix appearing in the left square, would play a part.

\subsection*{$\widetilde{P}_{\widetilde{G}G}$ - Projection as a subtractive process }

Recalling eq. \ref{eq:H-straight-sum}, $\mathcal{H}=\widetilde{\mathcal{H}}\oplus\bar{\mathcal{H}}$, we express the identity operator as a sum of projectors,
\begin{equation}
\mathcal{I}=\widetilde{\mathcal{P}}+\bar{\mathcal{P}}\label{eq:I_eq_P_plus_P},
\end{equation}
where, by analogy to $\widetilde{\mathcal{P}}$, we define $\bar{\mathcal{P}}:=\sum_{j=\widetilde{N}+1}^{N} \ketbra{\bar{g}_j}{\bar{b}_j}$. Using  $\ket{\bar{g}_j}=\ket{g_j}\,\,\forall j=\widetilde{N}+1\ldots N$,
\begin{equation}\label{eq:tildeP_subtractive}
\widetilde{\mathcal{P}}=\IdentOp - \sum_{j=\widetilde{N}+1}^{N} \ketbra{g_j}{\bar{b}_j}.
\end{equation}

We shall apply both sides of eq. \ref{eq:tildeP_subtractive} to $\ket{g_k}_{k\le\widetilde{N}}$. On the l.h.s, using eq. \ref{eq:P_Dirac} gives
\begin{equation}\label{eq:I_am_so_tired}
\widetilde{\mathcal{P}}\ket{g_k}=\ket{\widetilde{g}_k}.
\end{equation}
Applying the r.h.s of eq. \ref{eq:tildeP_subtractive} to $\ket{g_k}$ and equating it with eq. \ref{eq:I_am_so_tired} results in
\begin{equation}\label{eq:tildeP_subtractive_on_gk}
\ket{\widetilde{g}_k}=\ket{{g}_k} - \sum_{j=\widetilde{N}+1}^{N} \braket{g_j}{g_k} \ket{\bar{b}_j}\,\,\,\forall k\le\widetilde{N}.
\end{equation}

Note that $\widetilde{\mathcal{P}}$, applied to $\ket{g_k}_{k>\widetilde{N}}$ is zero, as the state is part of the basis spanning the complementary subspace.

We now have a new insight into the deformation of the Gaussians due to the basis reduction, as depicted in fig. \ref{fig:deformed_Gaussians}: it is the result of subtraction of $\ket{\bar{b}}$ vectors, weighted by the overlap of the Gaussians inside the reduced subspace with the Gaussians spanning the complementary space. This clarifies why Gaussians at the boundary of the reduced subspace are modified more --- they have a larger overlap with Gaussians outside the reduced subspace.

Applying $\widetilde{\mathcal{P}}$ to a superposition state $\ket{\psi}=\sum_{k=1}^{N}\gamma_k \ket{g_k}$ and using Eq.
\ref{eq:I_am_so_tired} gives
\begin{equation}\label{eq:tildeP_subtractive_on_psi}
\widetilde{\mathcal{P}}\ket{\psi}=\sum_{k=1}^{\widetilde{N}} \gamma_k \ket{\widetilde{g}_k}.
\end{equation}

Gaussians spanning the complementary subset are dropped, and the coefficients of the remaining Gaussians are unchanged but now serve as coefficients of the modified Gaussians. If this were an orthogonal basis, these coefficients would have been unchanged without modifying the underlying basis.

An equivalent view is provided by examining the projector representation $\widetilde{P}_{\widetilde{G}G}$. Define the matrix
\begin{equation}
R_{N\times\widetilde{N}}:=\left(\begin{array}{cccc}
1 & 0 & 0 & \ldots\\
0 & 1 & \ddots & \ddots\\
\vdots & \ddots & \ddots & 0\\
0 & \ldots & 0 & 1\\
0 & 0 & \ldots & 0\\
0 & \ddots &  & 0\\
\vdots &  & \ddots & \vdots\\
0 & 0 & \ldots & 0
\end{array}\right),\label{eq:R_def}
\end{equation}
allowing us to write
\begin{equation}
\widetilde{B}:=BR\text{.}\label{eq:B_red}
\end{equation}
Combining eqs. \ref{eq:P_in_FG}, \ref{eq:P_YX} and \ref{eq:G_tilde_def}, and using eqs. \ref{eq:B_red} and \ref{eq:bi-otrho prelim_B_first}, we obtain
\begin{equation}
\widetilde{P}_{\widetilde{G}G}=R^{\dagger}.\label{eq:P_GtildeG}
\end{equation}

\begin{figure}
\noindent \begin{centering}
\includegraphics[clip,trim=1cm 4cm 1cm 4cm,scale=0.4214]{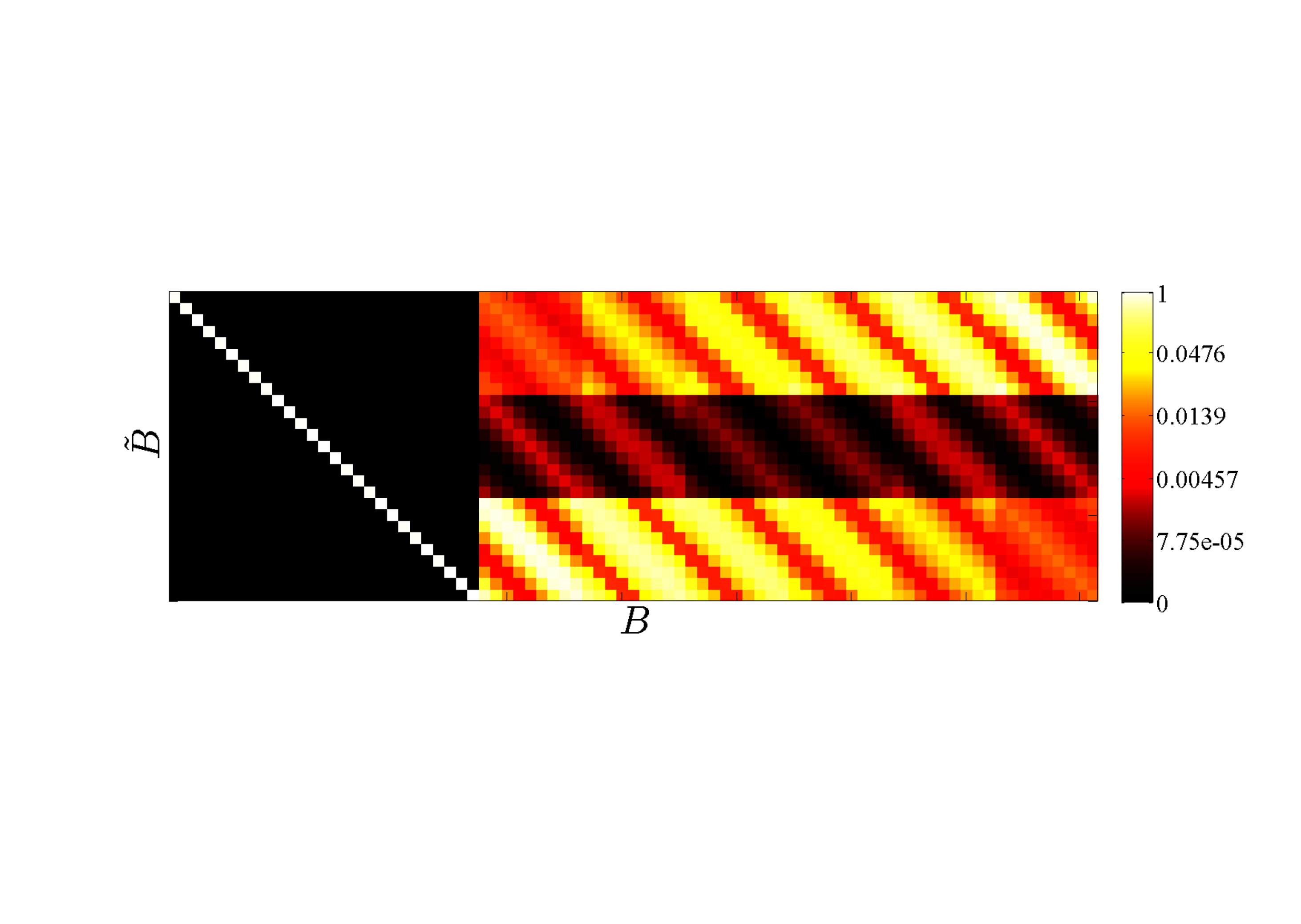}

\includegraphics[clip,trim=1cm 4cm 1cm 4cm,scale=0.4214]{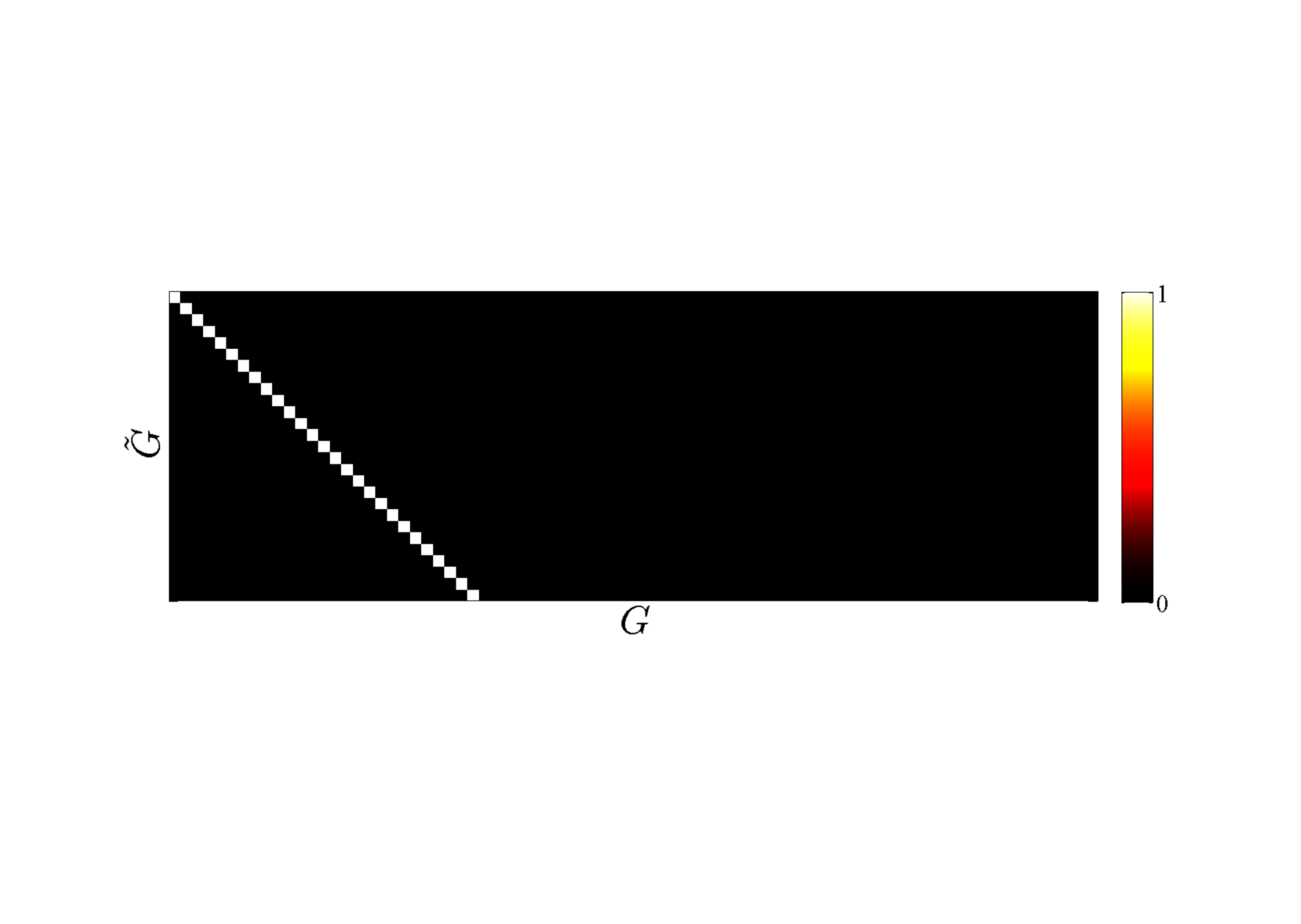}
\protect\caption{\label{fig:P_BB}(top panel) The $\widetilde{P}_{\widetilde{B}B}$ projector, i.e. the operator
projecting the state $\psi_{B}$, into the reduced Hilbert space $\widetilde{\mathcal{H}}$
and representing the result in the $\widetilde{B}$ basis. Notice the block
structure. The left square is the identity matrix, since
if $\protect\ket{\psi}=\protect\ket{b_{k\le\widetilde{N}}}$ it remains
unchanged by the projection. The right block provides insight into the unique additive contributions characteristic of non-orthogonal projections: as $\left\{ \protect\ket{b_{j}}\right\} _{j=1}^{\widetilde{N}}$
and $\left\{ \protect\ket{b_{j}}\right\} _{j=\widetilde{N}+1}^{N}$ overlap,
when projecting into $\widetilde{\mathcal{H}}$, all $b$ vectors have
some overlap into the reduced subspace, and therefore contribute to
the projected vector. (bottom panel) The $\widetilde{P}_{\widetilde{G}G}$ projector. The right block is empty, denoting the subtraction of all $\bar{G}$ vectors, which span the complementary subspace $\bar{\mathcal{H}}$, from the $G$ representation of the state. The left block is the identity matrix, denoting the coefficients in this block remain unchanged. However, since $\ket{\widetilde{g}}\neq\ket{g}$, the vector represented by these coefficient does change, due to the subtraction of the $\bar{G}$ subspace from the $\check{G}$ vectors, resulting in the $\widetilde{G}$ basis.}

\par\end{centering}

\noindent \centering{}
\end{figure}

This uniquely simply form is depicted in the bottom panel of fig. \ref{fig:P_BB}.

$\widetilde{P}_{\widetilde{B}B}$ and $\widetilde{P}_{\widetilde{G}G}$ provide two complementary views of the non-orthogonal projection. With $\widetilde{P}_{\widetilde{B}B}$, projection is an additive process: The vectors of the reduced basis, $\left\{\widetilde{b}_j\right\}_{j=1}^{ \widetilde{N} }$ are the same as their unreduced counterparts, $\left\{b_j\right\}_{j=1}^{\widetilde{N}}$, as is the case for orthogonal projections. However, the projection modifies their coefficients due to additive contributions by $\left\{b_j\right\}_{j=\widetilde{N}+1}^{N}$, which is unique to non-orthogonal projections. With $\widetilde{P}_{\widetilde{G}G}$, projection is a subtractive process: If $\ket{\psi}=\sum_{k=1}^{N}\gamma_k \ket{g_k}$, then the coefficients $\left\{\gamma_k\right\}_{k=1}^{\widetilde{N}}$ are unchanged by the projection and the rest are removed completely, as in an orthogonal projection. However, the Gaussian basis vectors are modified by subtraction proportional to their overlap with the Gaussians spanning the complementary subspace, changing $\ket{g}$ to $\ket{\widetilde{g}}$. This modification is again unique to non-orthogonal projections.

\begin{figure}
\noindent \begin{centering}
\includegraphics[scale=0.6173]{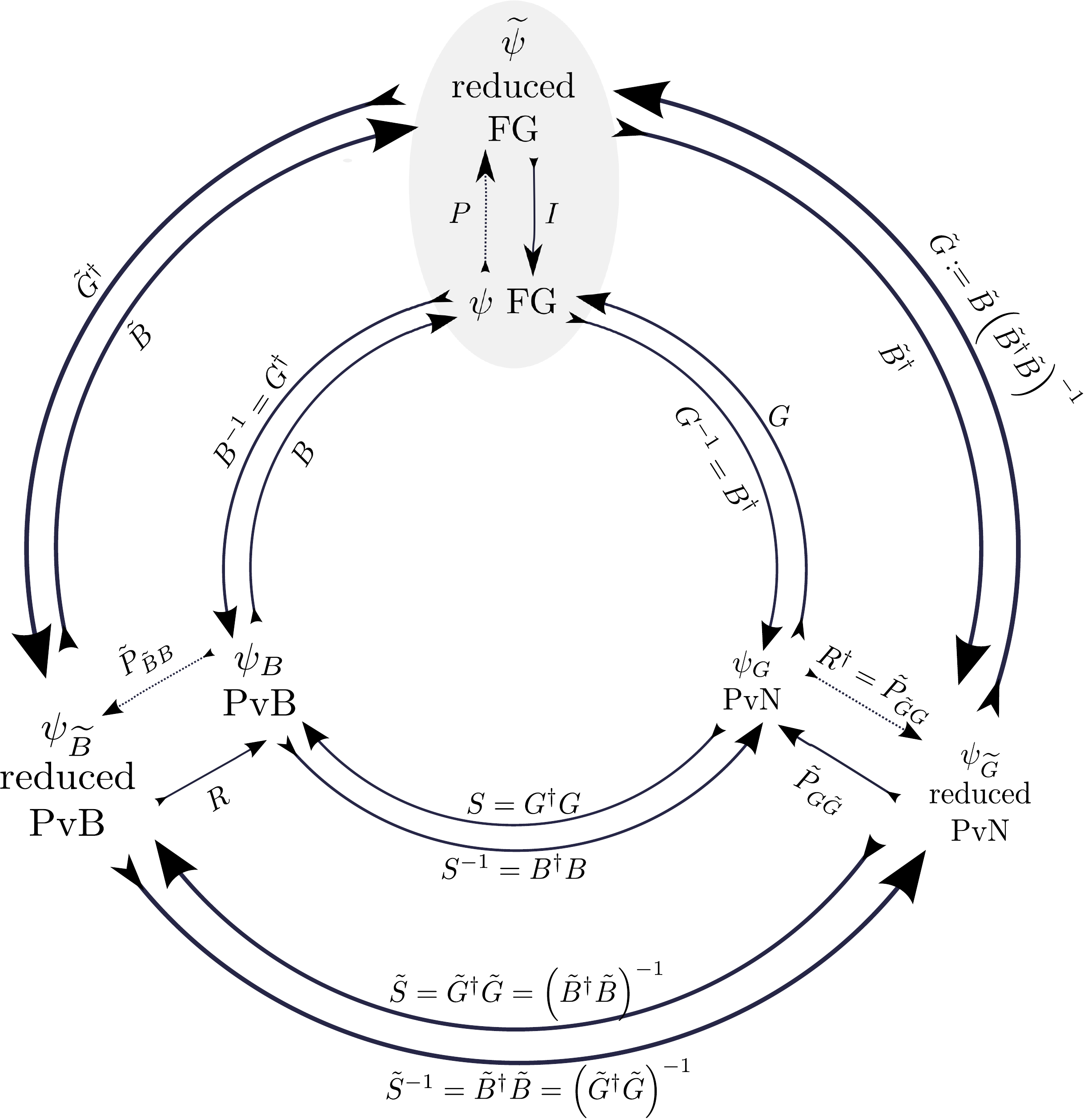}
\par\end{centering}

\noindent \centering{}\protect\caption{\label{fig:PEACE_sign} Transitions between the unreduced
and reduced bases, represented on the middle and outer circles,
respectively, with dimension $N$ and $\widetilde{N}$. Dashed arrows indicate
a projection (surjective; with information loss), continuous arrows
are injective (one-to-one but not necessarily on-to) mappings. The
nomenclature $\widetilde{P}_{YX}$ is used to indicate the projection
from representation $X$ of $\mathcal{H}$ into representation
$Y$ of $\widetilde{\mathcal{H}}$. This diagram may be used, for example, to construct $\widetilde{P}_{\widetilde{B}B}$ by
going through the following series of transformations: $\left(1\right)$, from the $B$ basis
to $G$ using the transformation matrix $S^{-1}$, $\left(2\right)$ projecting
with $R^{\dagger}$, $\left(3\right)$ transforming back to $\widetilde{B}$ using $\widetilde{S}$,
$\left(4\right)$ transforming to base $B$ using $R$. Multiplying the matrices appearing from right to left gives
$\widetilde{P}_{BB}=R\widetilde{S}R^{\dagger}S^{-1}$. }
\end{figure}

\begin{table}
\begin{changemargin}{-1.5cm}{-1.5cm}
\[
\begin{array}{cccccc}
\textrm{Hilbert space} & \textrm{dimension} & \textrm{b-like basis} & \textrm{g-like basis} & \textrm{Biorthogonality} & \textrm{Completeness}\\
\mathcal{H} & N & B={\left( G^{-1} \right)}^\dagger & G={\left( B^{-1} \right)}^\dagger & G^{\dagger}B={\IdentOp} & GB^{\dagger}={\IdentOp}\\
\widetilde{\mathcal{H}} & \widetilde{N} & \widetilde{B}=BR & \widetilde{G}:=\widetilde{B}\left(\widetilde{B}^{\dagger}\widetilde{B}\right)^{-1} & \widetilde{G}^{\dagger}\widetilde{B}={\IdentOp} & \left\{ \begin{array}{c}
\widetilde{G}=\widetilde{B}\left(\widetilde{B}^{\dagger}\widetilde{B}\right)^{-1}\\
\widetilde{B}=\widetilde{G}\left(\widetilde{G}^{\dagger}\widetilde{G}\right)^{-1}
\end{array}\right.\\
\bar{\mathcal{H}} & N-\widetilde{N} & \bar{B}:=\bar{G}\left(\bar{G}^{\dagger}\bar{G}\right)^{-1} & \bar{G}=G\bar{R} & \bar{G}^{\dagger}\bar{B}={\IdentOp} & \left\{ \begin{array}{c}
\bar{G}=\bar{B}\left(\bar{B}^{\dagger}\bar{B}\right)^{-1}\\
\bar{B}=\bar{G}\left(\bar{G}^{\dagger}\bar{G}\right)^{-1}
\end{array}\right.
\end{array}
\]
\protect\caption{\label{tab:Summary-of-attributes}Summary of properties of the full, reduced and complementary Hilbert spaces.}
\end{changemargin}
\end{table}

\subsection{\label{sub:check-check}Projecting into the subspace spanned by Gaussians,
$\check{\mathcal{H}}$}

We have defined the reduced subspace $\widetilde{\mathcal{H}}$ in eq.
\ref{eq:red_subspace_def} as the Hilbert space spanned by $\left\{ \ket{b_{k}}\right\} _{k=1}^{\widetilde{N}}$.
One may define a different reduced subspace,
$\check{\mathcal{H}}=\textrm{span}\left(\left\{ \ket{g_{k}}\right\} _{k=1}^{\widetilde{N}}\right)$,
as in eq. \ref{eq:check_Hilbert}. As discussed in Section \ref{sub:Bi-orthogonal-bases-for}, these subspaces are not identical (unless $\widetilde{N}=N$), as is evident when $\widetilde{N}=1$ and $\widetilde{\mathcal{H}}=\textrm{span}\left(\ket{b_1}\right)$ and $\check{\mathcal{H}}=\textrm{span}\left(\ket{g_1}\right)$.

Following the discussion in Sections
\ref{sub:Definition_of_reduced}, \ref{sub:Bi-orthogonal-bases-for}
and \ref{sub:The-Projector} one may define the projector
into this subspace and the subspace's biorthogonal bases. Let us denote the
first $\widetilde{N}$ columns of the $G$ matrix with the matrix $\check{G}_{N\times\widetilde{N}}$,
i.e. $\check{G}$ is the $\Theta$ representation of $\left\{\ket{g_{k}}\right\}_{k=1}^{\widetilde{N}}$, the basis spanning
$\check{\mathcal{H}}$. To satisfy the biorthogonality requirement,
eq. \ref{eq:bi-otrho prelim_G_first}, in the $\check{\mathcal{H}}$ basis,
one must define

\begin{equation}
\check{G}:=GR\label{eq:G_check}
\end{equation}

\begin{equation}
\check{B}:=\check{G}\check{S}^{-1}.\label{eq:B_check}
\end{equation}

with $\check{S}=\check{G}^{\dagger}\check{G}$. The projector into
$\check{\mathcal{H}}$ may then be defined as

\begin{equation}
\check{\mathcal{P}}=\sum_{j=1}^{\widetilde{N}}\ketbra{\check{b}_{j}}{\check{g}_{j}},
\end{equation}

with $\ket{\check{g}_{j}}=\ket{g_{j}}$. Therefore, in the $\Theta$
representation the projector is
\begin{equation}
\check{P}=\check{G}\check{B}^{\dagger}\label{eq:P_check}
\end{equation}

Note both $\bar{\mathcal{H}}$ and $\check{\mathcal{H}}$ are spanned by Gaussians. Specifically, $\bar{\mathcal{H}}$ is the reduced subspace spanned by the set of Gaussians complementary to $\check{G}$. It must be stressed that $\widetilde{\mathcal{H}}\neq\check{\mathcal{H}}\neq\bar{\mathcal{H}}$.

As discussed in Section \ref{sub:PvN,PvB}, the representation of a state as a sum of of Gaussians (PvN) is not sparse.
Therefore, given a well-localized state $\ket{\psi}$,
if one wishes to use the $\check{\mathcal{H}}$ subspace, one can expect that a very large number of functions will be needed.
This is in stark contrast to $\mathcal{\widetilde{H}}$, where well-localized states are well-approximated by a small number of functions. Thus, given a fixed size of the representation, one will achieve much higher accuracies
with $\mathcal{\widetilde{P}}$ than with $\check{\mathcal{P}}$. This
will be explicitly demonstrated when we discuss the various representations
of the reduced Hamiltonian and their accuracy, in section \ref{sub:The-TDSE-for-the-reduced_state}.

\section{\label{sub:The-TDSE-for-the-reduced_state}The
Hamiltonian and the Schr{\"o}dinger equation for the reduced state}

The Schr{\"o}dinger equation for the reduced state requires projecting the Hamiltonian into the reduced subspace. In this section we consider three possibilities from both the theoretical and practical viewpoints: Projecting
into the $\widetilde{\mathcal{H}}$ subspace (Section \ref{sub:H_BB}),
into the $\check{\mathcal{H}}$ subspace (Section \ref{sub:H_check_check})
and performing a pseudo-projection into a combination of the two (Section
\ref{sub:H_Hybrid}). Finally we benchmark the respective accuracies of the possible projections (Section \ref{sub:H_Comparing}).

\subsection{\label{sub:H_BB}The equation of motion in the $\widetilde{\mathcal{H}}$ subspace}

To arrive at the reduced Hamiltonian, we apply the $\widetilde{P}$ projection
on both the input and output of the Hamiltonian. As the state is represented
in the $\widetilde{B}$ basis, one can either convert the vector to $\Theta$,
project into $\widetilde{\mathcal{H}}$, apply $H$, project the resulting
state again into $\widetilde{\mathcal{H}}$, and transform it to $\widetilde{B}$,
or, equivalently, first apply the projection and then the base transformation.
These two alternatives correspond to the following two expressions for $H_1$:

\begin{equation}
H_{1}:=\widetilde{G}^{\dagger}\left(\widetilde{P}H\widetilde{P}\right)\widetilde{B}=\widetilde{P}_{\widetilde{B}B}\left(G^{\dagger}HB\right)\widetilde{P}_{B\widetilde{B}}.\label{eq:H1_def}
\end{equation}

Utilizing eq. \ref{eq:P_in_FG}, \ref{eq:P_BB_1} and \ref{eq:G_tilde_def}
\begin{equation}
H_{1}=\widetilde{G}^{\dagger}H\widetilde{B}=\left(\widetilde{B}^{\dagger}\widetilde{B}\right)^{-1}\left(\widetilde{B}^{\dagger}H\widetilde{B}\right).
\label{eq:H_tilde_B_tilde_B}\end{equation}

The TISE and TDSE then take the form

\begin{equation}
E\psi_{\widetilde{B}}=H_{1}\psi_{\widetilde{B}}\label{eq:TISE_red}
\end{equation}
\begin{equation}
\partial_{t}\psi_{\widetilde{B}}=-\frac{i}{\hbar}H_{1}\psi_{\widetilde{B}}.\label{eq:TDSE_red}
\end{equation}

\emph{Eq. \ref{eq:H_tilde_B_tilde_B}, \ref{eq:TISE_red} and \ref{eq:TDSE_red} will be the key working equations for the remainder of this article.}

The formulation using only $\widetilde{B}$ matrices (r.h.s. of \ref{eq:H_tilde_B_tilde_B})
is appealing, as $\widetilde{B}=BR$ is the naturally reduced matrix
in $\widetilde{P}$. An equivalent, more heuristic approach, is to replace
all matrices in any of the forms of eq. \ref{eq:H_GB_BB_BG_GG}
by their reduced counterparts, $B\longrightarrow\widetilde{B}$,
$G\longrightarrow\widetilde{G}$, $S\longrightarrow\widetilde{S}$, etc. All four variations transform to the reduced Hamiltonian in eq. \ref{eq:H_tilde_B_tilde_B}.

Finally, we note that $H_{1}$ is similar to a Hermitian matrix, in that the product of matrices to the left and right of $H$ in eq. \ref{eq:H_tilde_B_tilde_B} produce the identity, $\left(\left(\widetilde{B}^{\dagger}\widetilde{B}\right)^{-1}\widetilde{B}^{\dagger}\right)\left(\widetilde{B}\right)={\IdentOp}$. Therefore eigenvalues are real and evolution is unitary. Moreover, combining eq. \ref{eq:H_tilde_B_tilde_B} and \ref{eq:TISE_red} the TISE may be reformulated as a generalized eigenvalue equation with Hermitian matrices.

\begin{equation}
\left(\widetilde{B}^{\dagger}\widetilde{B}\right)\lambda\psi_{\widetilde{B}}=\left(\widetilde{B}^{\dagger}H\widetilde{B}\right)\psi_{B}\label{eq:Scrod_eig_red}
\end{equation}
thus avoiding the matrix inversion required to compute $\widetilde{S}$.
This form is solvable with an iterative eigensolver, such as Arnoldi.
Unfortunately, in this form it is not possible to implement the eigensolver with just matrix-vector multiplications (i.e. it is not possible to avoid matrix-matrix multiplications) \cite{Book-Templates-Solution-Eigenvalue-Problems}. This is in contrast with the pseudo-projection discussed in Section \ref{sub:H_Hybrid}, which can be implemented with just matrix-vector multiplications.

While accurate and mathematically rigorous, computing the $H_{1}$
form in eq. \ref{eq:H_tilde_B_tilde_B} is time consuming. Most significantly,
calculation of elements of the reduced Hamiltonian $\widetilde{B}^{\dagger}H\widetilde{B}$
can be laborious, particularly in the multi-dimensional case, as the
$B$ functions are non-localized. Fortunately, there are symmetry
considerations and numerical techniques which can accelerate this
computation by several orders of magnitude. Alternatively,
using eq. \ref{eq:H_tilde_B_tilde_B}, \ref{eq:B_red}, \ref{eq:Overlap-matrices} and \ref{eq:Overlap-matrices-red}, one may rewrite $H_{1}$ as

\begin{equation}
H_{1}=\widetilde{S}\left(R^{\dagger}S^{-1}G^{\dagger}HGS^{-1}R\right)\label{eq:H1_on_the_fly}
\end{equation}

Eq. \ref{eq:H1_on_the_fly} may be quicker to compute then \ref{eq:H1_def}: due to the locality of $G$ and
the fact that elements of $H$ are usually functions of either position or
momentum (but rarely both), the vast majority of $G^{\dagger}HG$
elements are vanishingly small. Note that $G$ and $S^{-1}$ decompose dimensionally.
Indeed, one would still have to calculate $\widetilde{S}$, which requires
the inversion of $\widetilde{B}^{\dagger}\widetilde{B}$, but this may be
accelerated, if one is able to store the matrix. If we then use an iterative algorithm
for solving the TISE (e.g. Arnoldi) and the TDSE (e.g. Taylor propagator), one may use only matrix-vector
type operations to further accelerate the process. Similar considerations have been noted in \cite{TuckerCarrington}.

\subsection{\label{sub:H_check_check}Alternatives to the $\widetilde{\mathcal{H}}$-projected equations of motion}

In the previous section we presented the projection of the TISE and TDSE into the $\widetilde{\mathcal{H}}$ subspace. There are, however, additional alternatives we should consider. First, we shall examine the possible choices for state representation: In addition to the $\widetilde{B}$ representation which we have been using, there are have three additional a priori candidates: $\widetilde{G}$, $\check{B}$ and $\check{G}$. Then, we shall consider various projections for the Hamiltonians. This analysis is of some importance, as alternatives to eq. \ref{eq:H_tilde_B_tilde_B} are currently in use by the community. For example, $H_{2}$, defined in eq. \ref{eq:H2} using a hybrid of $\widetilde{P}$ and $\check{P}$ is used in \cite{TuckerCarrington}.

The first alternative state representation, $\widetilde{G}$, spans $\widetilde{\mathcal{H}}$ just as $\widetilde{B}$ does and therefore provides the same accuracy. However, it has a disadvantage: the coefficients are determined by an overlap with $\widetilde{B}$-s, which are highly non-local. As a result, the coefficients are not expected to drop-off sharply at the boundary of the reduced subspace, where the wavepacket is absent. As a result, we cannot use their declining values to determine the selection of basis functions for the reduced subspace, as one may do with an amplitude cutoff threshold for $\psi_{\widetilde{B}}$. This issue is of primary concern, especially for the TDSE, where the coefficients of the state at the boundary of the reduced phase space serves determine how to modify the subspace over time. We therefore conclude that the $\widetilde{G}$ is unfit for our purposes.

The two remaining options, $\check{B}$ and $\check{G}$, are disqualified because the subspace to which such states belong, $\check{H}$, is spanned by Gaussians, and hence does not provide a sparse representation for localized states. In other words, many Gaussians (or many $\check{B}$ vectors) are required to accurately represent a localized state (see Section \ref{sub:check-check}). We therefore conclude that $\widetilde{B}$ is the only viable option for state representation.

Next we proceed to examine the possible projections of the Hamiltonian. Until now we have been considering projection into $\widetilde{\mathcal{H}}$. The other natural alternative is projection into $\check{\mathcal{H}}$.
This may be performed either by  acting with $\check{\mathcal{P}}$ on $H$ on the left and the right, or by using the replacement rules $G\longrightarrow\check{G}$, $S\longrightarrow\check{S}=\check{G}^{\dagger}\check{G}$, $B\longrightarrow\check{B}=\check{G}\check{S}^{-1}$ on any of the Hamiltonian forms in eq. \ref{eq:H_GB_BB_BG_GG}. This leads to

\begin{equation}
H_{4}:=\check{G}^{\dagger}\check{P}H\check{P}\check{B}=\check{G}^{\dagger}H\check{B}=\left(\check{G}^{\dagger}H\check{G}\right)\left(\check{G}^{\dagger}\check{G}\right)^{-1}.\label{eq:H4}
\end{equation}

Note that we would like to continue to represent states in the $\widetilde{B}$ representation, in accordance with the above discussion of its compactness. Then the TISE and TDSE take the form

\begin{equation}
E\psi_{\widetilde{B}}=H_{4}\psi_{\widetilde{B}}\label{eq:TISE_red_H4}
\end{equation}
\begin{equation}
\partial_{t}\psi_{\widetilde{B}}=-\frac{i}{\hbar}H_{4}\psi_{\widetilde{B}}.\label{eq:TDSE_red_H4}
\end{equation}

However, because the subspace into which $H_{4}$ is projected does not span the subspace defined by the  $\psi_{\widetilde{B}}$, the accuracy may be significantly lower than that of $H_{1}$ (see fig. \ref{fig:H1v2v3v4} and the discussion in Section \ref{sub:H_Comparing}).
Nevertheless, $H_{4}$ greatly benefits from having a highly local Hamiltonian transform, $\check{G}^{\dagger}H\check{G}$, making it usable when fast low-accuracy results are acceptable.

With the strict application of $\widetilde{P}$ leading to an accurate but slow $H_{1}$, and the application of $\check{P}$ leading
to an inaccurate alternative, in the following section we explore the possibility of a hybrid projection that combines some of the benefits of both.

\subsection{\label{sub:H_Hybrid}The equation of motion in hybrid subspaces}

Recall the discussion in Section \ref{sub:Bi-orthogonal-bases-for} regarding the modification of the $\widetilde{G}$
Gaussians when near the edge of the reduced subspace. One may hypothesize that if
we extend the reduced subspace beyond the minimal volume required
to represent the state, so that Gaussians near the boundary do not have significant overlap with the state,
we may achieve a good approximation while still using the unmodified
Gaussians. In the nomenclature established in Section \ref{sub:check-check},
we wish to replace $\widetilde{G}$ with $\check{G}$, and possibly avoid
the non-local $\widetilde{B}$ matrix altogether.

We note that $\check{G}^{\dagger}\widetilde{B}={\IdentOp}_{\widetilde{N}}$, i.e. they
are biorthogonal. However, as they do not span the same subspace the completeness relations do not hold.

Recalling eq. \ref{eq:H_tilde_B_tilde_B}, $H_{1}=\widetilde{G}^{\dagger}H\widetilde{B}$
and eq. \ref{eq:H4}, $H_{4}=\check{G}^{\dagger}H\check{B}$, we shall define the hybrid form

\begin{equation}
\begin{array}{c}
H_{2}:=\check{G}^{\dagger}H\widetilde{B},\end{array}\label{eq:H2}
\end{equation}

with the corresponding TISE and TDSE:

\begin{equation}
E\psi_{\widetilde{B}}=H_{2}\psi_{\widetilde{B}}\label{eq:TISE_red1}
\end{equation}
\begin{equation}
\partial_{t}\psi_{\widetilde{B}}=-\frac{i}{\hbar}H_{2}\psi_{\widetilde{B}}.\label{eq:TDSE_red1}
\end{equation}

This form is advantageous over eq. \ref{eq:H_tilde_B_tilde_B} in
that no matrix inversion is necessary (implying there is no generalized
eigenvalue problem to solve for the TISE), and the localized nature
of $\check{G}$ can be used to simplify the calculation of multi-dimensional
terms in $\check{G}^{\dagger}H\widetilde{B}$. $H_{2}$ is similar to
a Hermitian matrix (as $\check{G}^{\dagger}\widetilde{B}={\IdentOp}_{\widetilde{N}}$)
and hence diagonalizable with real eigenvalues.

Eq. \ref{eq:H2} may be reformulated using eq. \ref{eq:G_check}, \ref{eq:B_red} and \ref{eq:Overlap-matrices} as

\begin{equation}\label{eq:H2}
H_{2}=R^{\dagger}G^{\dagger}HGS^{-1}R.
\end{equation}

This form allows the TDSE to be solved using only matrix-vector type operations. $S^{-1}$ decomposes into a product of one-dimensional matrixes and $G^{\dagger}HG$ may be calculated analytically by expanding the potential term as a Taylor series around the center of the Gaussian. Such an approach may be sufficiently quick to eliminate the need to store the reduced Hamiltonian, by making use of an iterative eigensolver as described in \cite{TuckerCarrington} along with a Taylor propagator for the TDSE.

A final possible hybrid is $H_{3}:=\widetilde{G}^{\dagger}H\check{B}$ but we do not see any immediate usefulness in this form.

\subsection{\label{sub:H_Comparing}Comparing the accuracy of the dynamical formulations}

\begin{figure}
\noindent \begin{centering}
\includegraphics[scale=0.3869]{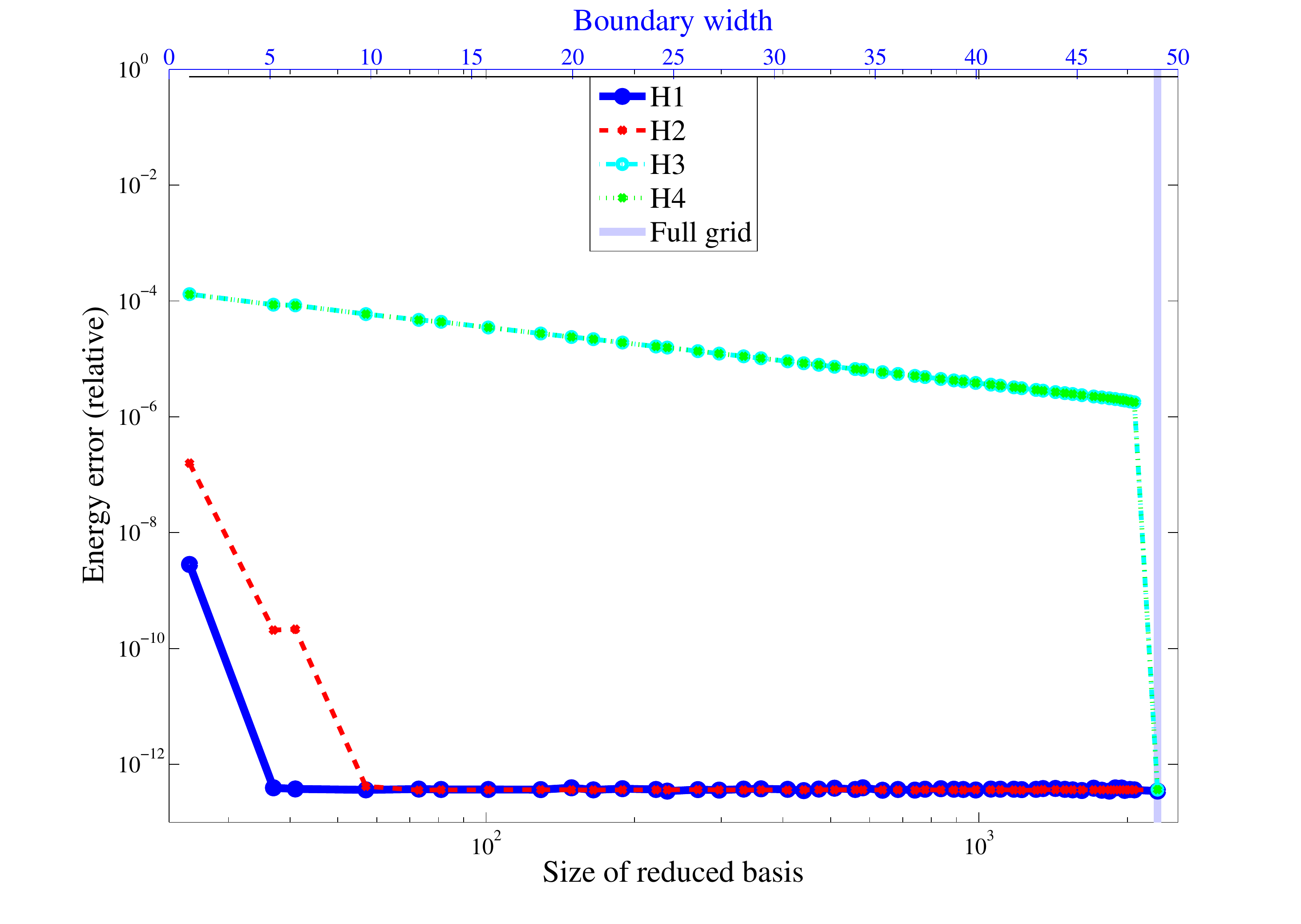}
\par\end{centering}

\noindent \centering{}\protect\caption{\label{fig:H1v2v3v4}Accuracy of the ground state energy of the 1D Harmonic
oscillator as a function of the reduced basis size (lower $x$-axis) or reduced-basis boundary width (upper $x$-axis) computed using different reduced Hamiltonians. As expected from the theoretical discussion, $H_{1}=\widetilde{G}^\dagger H \widetilde{B}$
is orders of magnitude more accurate then $H_{2}=\check{G}^\dagger H \widetilde{B}$, until the reduced subspace is increased to the point where the state does not significantly overlap any of the distorted Gaussians, and the distinction between $\widetilde{G}$ and $\check{G}$ becomes moot. $H_3=\widetilde{G}^\dagger H \check{B}$ and $H_{4}=\check{G}^\dagger H \check{B}$ are significantly less accurate.}
\end{figure}

We now compare the accuracy of the various reduced Hamiltonians introduced
in the previous sections. From fig. \ref{fig:H1v2v3v4} we can conclude that $H_{1}$ is indeed the most accurate. This is expected as it contains no approximations beyond the amplitude cutoff used to define the reduced subspace. $H_{2}$ provides
two orders of magnitude lower accuracy for the same reduced basis
size or alternatively, requires a boundary width of $5$ lattice cells
to achieve equivalent accuracy. The latter translates to $30\%$
to $80\%$ larger reduced basis size for 1D problems, and is expected
to grow exponentially with higher dimensions, making it an unappealing
option for high-dimensional problems. The lower accuracy of $H_2$ may be understood when we compare equations \ref{eq:H2} to \ref{eq:H_tilde_B_tilde_B}, i.e. $\check{G}^{\dagger}H\widetilde{B}$ to $\widetilde{G}^{\dagger}H\widetilde{B}$, and recall the deformation of the Gaussians by the basis reduction, as depicted in fig. \ref{fig:deformed_Gaussians}. $H_2$ ignores this deformation, instead acting with the unmodified Gaussians. If we enlarge the boundary of the reduced subspace sufficiently so that there is little overlap between the state represented and the deformed Gaussians, then $\widetilde{G}^\dagger\psi \approxeq \check{G}^\dagger\psi$. Therefore, $H_2$ may achieve accuracy equal to $H_1$ provided the reduced subspace is expanded.

$H_{3}$ and $H_{4}$ provide almost identically poor performance, never achieving an accuracy of $10^{-5}$, except when
utilizing the full grid, where the difference between the various
reduced Hamiltonians disappears. This is again expected, as by using
the projector $\check{P}$ instead of $\widetilde{P}$ on the input state to the Hamiltonian, we are projecting
into a subspace spanned by Gaussians (Section \ref{sub:check-check}), which is non-sparse.
As analyzed in \cite{Asaf-PRL}, in this case the full Hilbert space is required to achieve highly accurate results.

To conclude, for highly accurate results, $H_{1}$ is the form of
choice. In some instances where high accuracy is not paramount, or
one opts for a significant increase in the reduced subspace size, $H_{2}$ may allow for greater speed.

\section{\label{sec:Examples}Examples}

We provide some illustrative examples of the use of the PvB representation for both the TISE and TDSE. The algorithms used to solve the time dependent and time independent Schr{\"o}dinger are described in detail in \cite{The-Alg-Paper} and an application of PvB to the challenging problem of double ionization of helium, is given in \cite{PvB-MCTDH-comparison-paper}.

\subsection{1D Morse Oscillator}
\label{sec:ex_morse}
\begin{figure}
 \noindent \begin{centering}
\includegraphics[width=.6\textwidth]{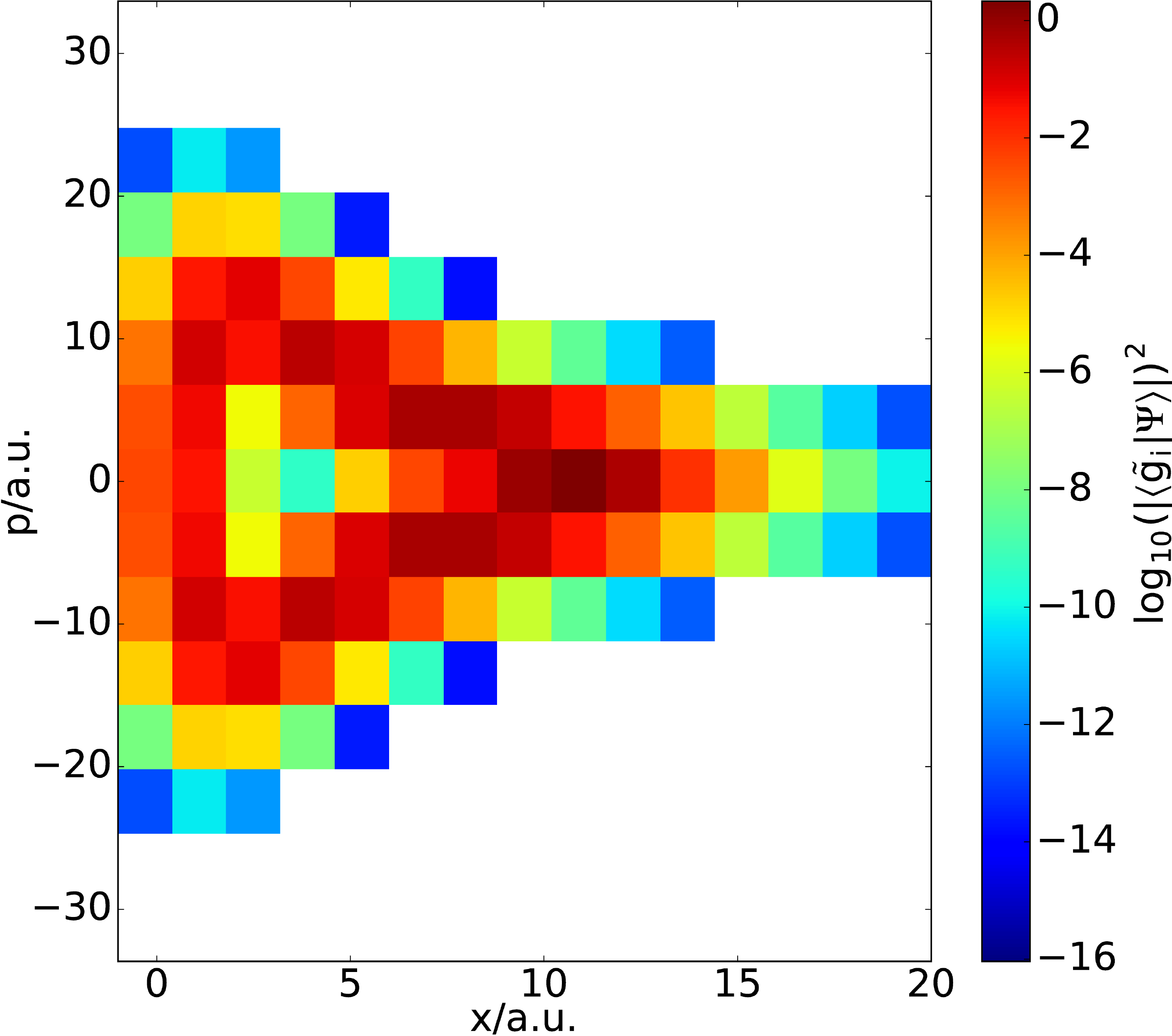}
\par\end{centering}
\noindent \centering{}\protect
\caption{PvB representation of the $21^\textrm{st}$ eigenstate of the one dimensional Morse oscillator, eq.~\eqref{eq:ex_morse}. Pixel colors represent the absolute value squared of the PvB expansion coefficients. Only the basis functions corresponding to colored pixels are part of the reduced subspace.\label{fig:ex_states_morse}}
\end{figure}

The Hamiltonian for the Morse oscillator, in atomic units, is
\begin{equation}
  H_\text{Morse} = -\frac{\partial_x^2}{2\times 6} + 12 \left( 1 - e^{-\frac{1}{2}\left({x-2}\right)} \right)^2.\label{eq:ex_morse}
\end{equation}
The PvB representation for the $21^\textrm{st}$ eigenstate is shown in fig.~\ref{fig:ex_states_morse}.

\begin{figure}
\centering
\includegraphics{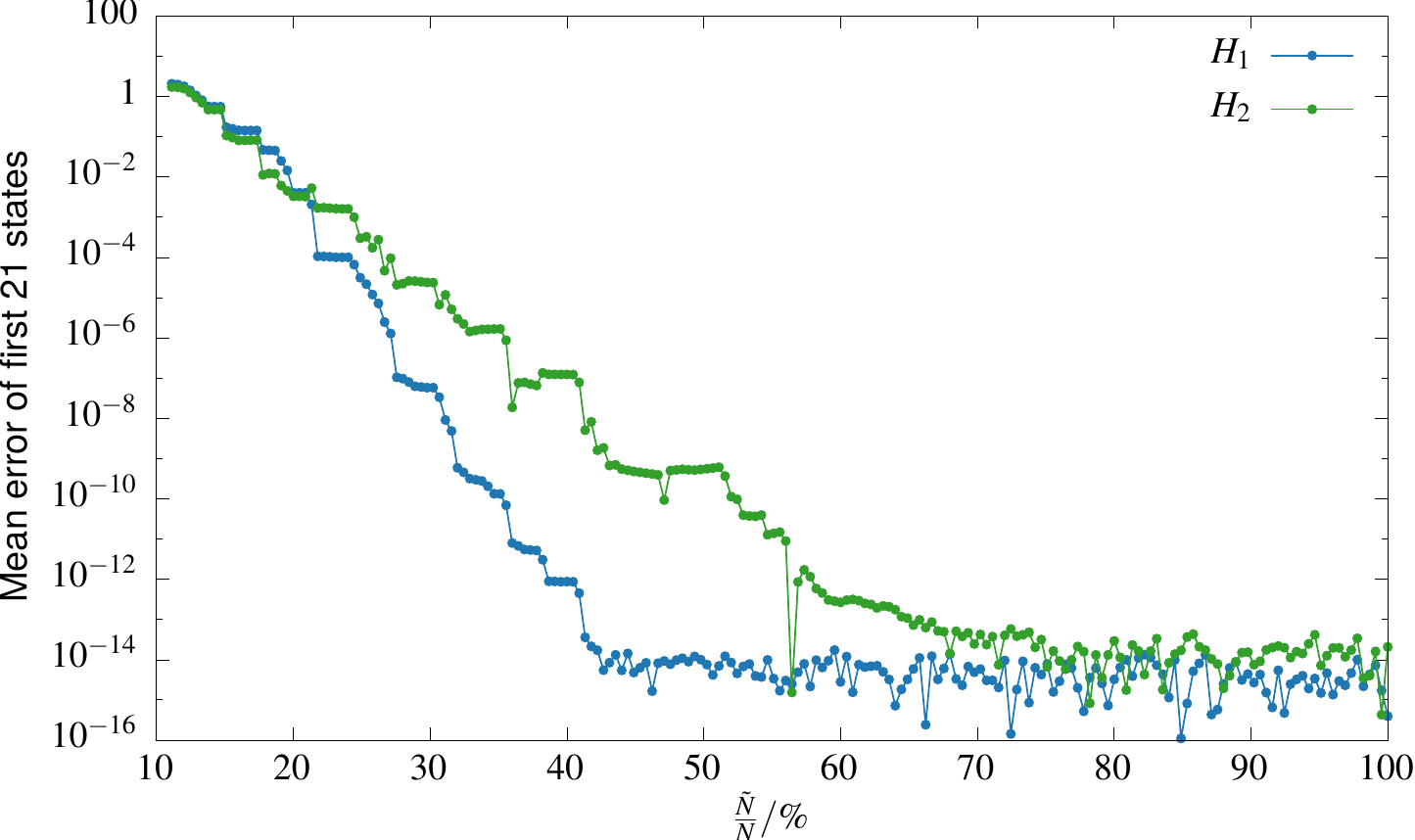}
\caption{Mean error of the first $21$ eigenstates in a one dimensional Morse oscillator, as a function of the ratio between the reduced and unreduced basis sizes. The system supports $24$ bound states; the size of the reduced basis is $97$. The error is shown for the two Hamiltonians, $H_1$ and $H_2$. Eigenstate error is computed as the infidelity with respect to the eigenstate computed with the unreduced Hilbert space.}
\label{fig:ex_inv_comp_morse}
\end{figure}

The mean error of the first $21$ states as a function of the reduced basis size is depicted in fig.~\ref{fig:ex_inv_comp_morse}. The behavior of $H_2$ compared to $H_1$ for the Morse oscillator is similar to their behavior for the harmonic oscillator ground state (fig. \ref{fig:H1v2v3v4}). If the reduced basis is large enough, so that the basis functions near the subspace boundary have negligible overlap with the state, then $\tilde{G}^\dagger \psi$ is almost equal to $\check{G}\psi$ and the results using $H_2$ do not differ significantly from the results using $H_1$. In other words, $\tilde{S}_{\tilde{N}\times\tilde{N}}=(\tilde B^\dagger \tilde B)^{-1}$ is similar to $\check{S} = R^T S_{N\times N}R$. At the other extreme, if the Hilbert space is pruned too drastically, parts of phase space with significant population are left out of the reduced subspace. In that case, the errors induced by the excessive pruning become more significant than the errors caused by approximating $\tilde{G}$ by $\check{G}$ and the distinction between $H_1$ and $H_2$ becomes negligible. We conclude that $H_2$ may be useful for low accuracy applications.

\subsection{2D Coupled Harmonic Oscillator}

\begin{figure}
\centering
\includegraphics{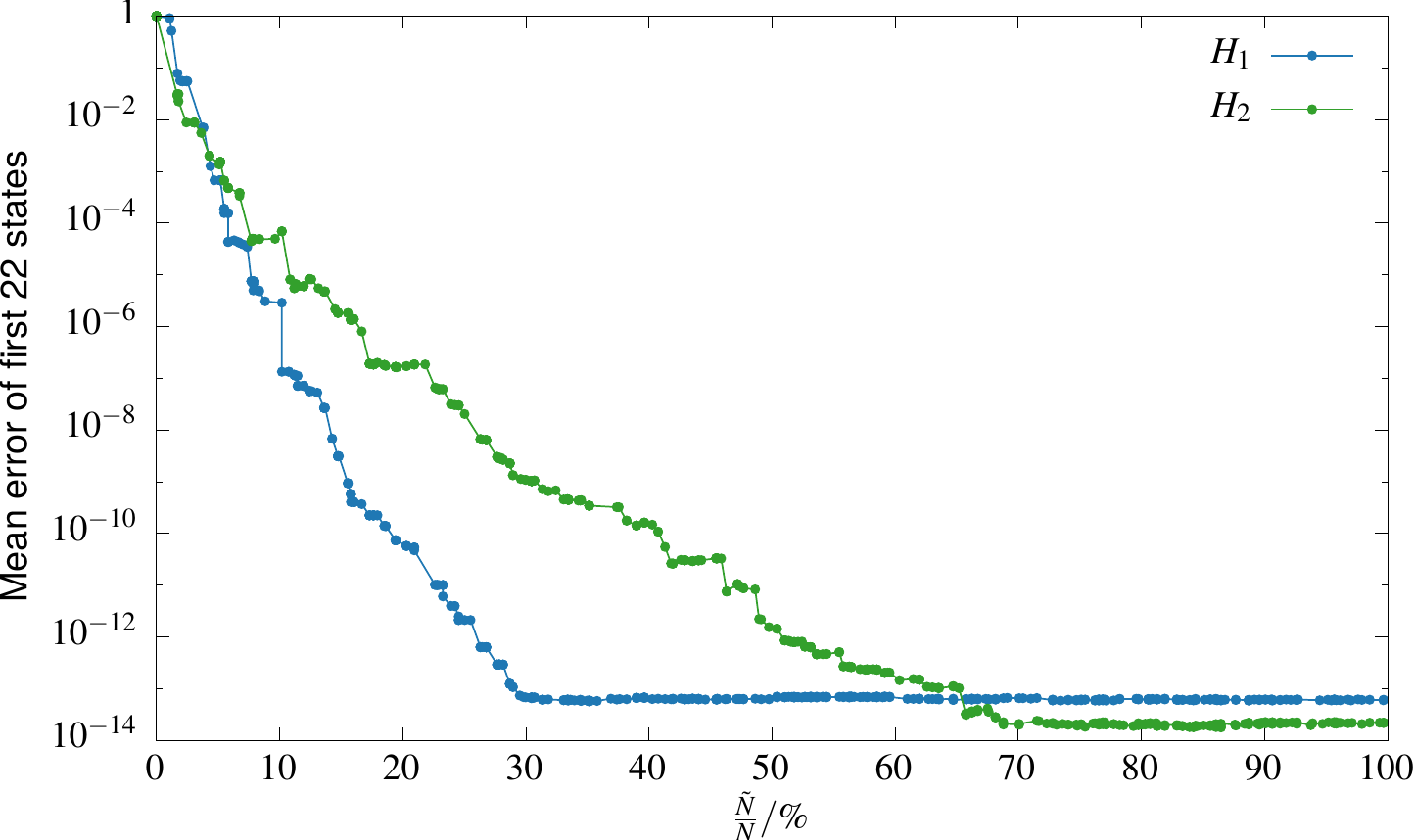}
\caption{Mean error of the first $22$ eigenstates in a two dimensional coupled harmonic oscillator, eq.~\eqref{eq:ex_2dho}, as a function of the ratio of the reduced and unreduced basis sizes. The error is shown for the two Hamiltonians, $H_1$ and $H_2$.}
\label{fig:ex_inv_comp_cHO2D}
\end{figure}

The observations of Section \ref{sec:ex_morse} are applicable to higher dimensional systems. We repeat the previous analysis for a coupled two-dimensional harmonic oscillator described by
\begin{equation}
  H_\text{2D-HO} = -\frac12(\partial_x^2+\partial_y^2) + \frac12 (x^2 + y^2) - 0.3 xy.\label{eq:ex_2dho}
\end{equation}
Figure~\ref{fig:ex_inv_comp_cHO2D} shows the error of the first $22$ eigenstates of this system as a function of the subspace size. The results are qualitatively similar to those of fig.~\ref{fig:ex_inv_comp_morse}. For a large reduced space, the results of $H_1$ and $H_2$ are similar, with the results of $H_2$ actually slightly more accurate. The slightly higher accuracy of $H_2$ may originate from the additional inversion of $\tilde S^{-1}$ that is required for $H_1$. For intermediate basis sizes, the error of $H_2$ is larger than for $H_1$, until the error introduced by excessive pruning of the basis dominates at small basis sizes.

\subsection{2D Double Well Dynamics}

\begin{figure}
\noindent \begin{centering}
\includegraphics[width=\textwidth]{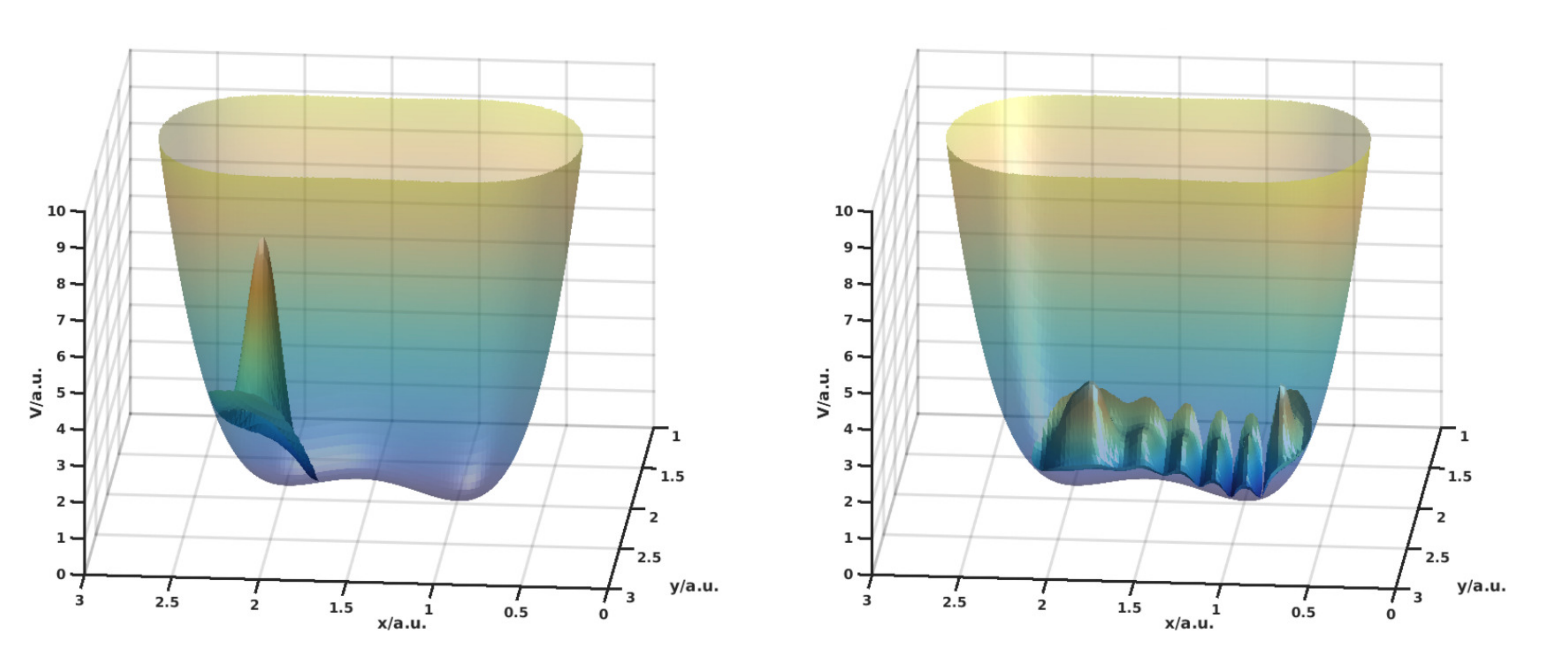}
\par\end{centering}
\noindent \centering{}\protect
\caption{Potential surface for 2D double well (eq.~\eqref{eq:ex_dwell}), overlaid with the initial (left panel) and final (right panel) state. Wavefunction $z$ coordinate is proportional to the absolute value squared of its amplitude.\label{fig:ex_dwell_states}}
\end{figure}

We now turn to the time-dependent case. When propagating a wavefunction in time, the reduced basis changes as the phase space occupied by the state evolves. To compare the accuracy of $H_1$ and $H_2$, we consider a two dimensional double well potential with strong coupling between $x$ and $y$, with
\begin{equation}
  H_\text{2D-DW} = -\frac1{2\times 200}(\partial_x^2+\partial_y^2) + 6.4 (x-1)^2 (x-2)^2 + 37.5 (y-2)^2 + 10  x^2 y,\label{eq:ex_dwell}
\end{equation}
and the initial wave packet is given by
\begin{equation}
 \braket{xy}{\psi_{\textrm{initial}}} = \frac{\sqrt{\frac{2}{\pi}}}{\left(0.04\times0.02\right)^{\frac{1}{4}}}\exp{\left(-{\frac{{\left( x - 2.1 \right)}^2}{0.04} - \frac{\left(y-2.05\right)^2}{0.02}}\right)}.
 \label{eq:ex_dwell_psiInitial}
\end{equation}
The wave packet was propagated from $t_0=0$ to $t_2=24.6$. The potential surface and the initial and final wave packets are shown in fig.~\ref{fig:ex_dwell_states}. At the final time, the wave packet has spread across the barrier and oscillates in $x$ and $y$. The wave packet was also investigated at $t_1=16.6$ (not shown), when the packet is at the barrier but still retains a compact form.

\begin{figure}
\centering
\includegraphics[width=\textwidth]{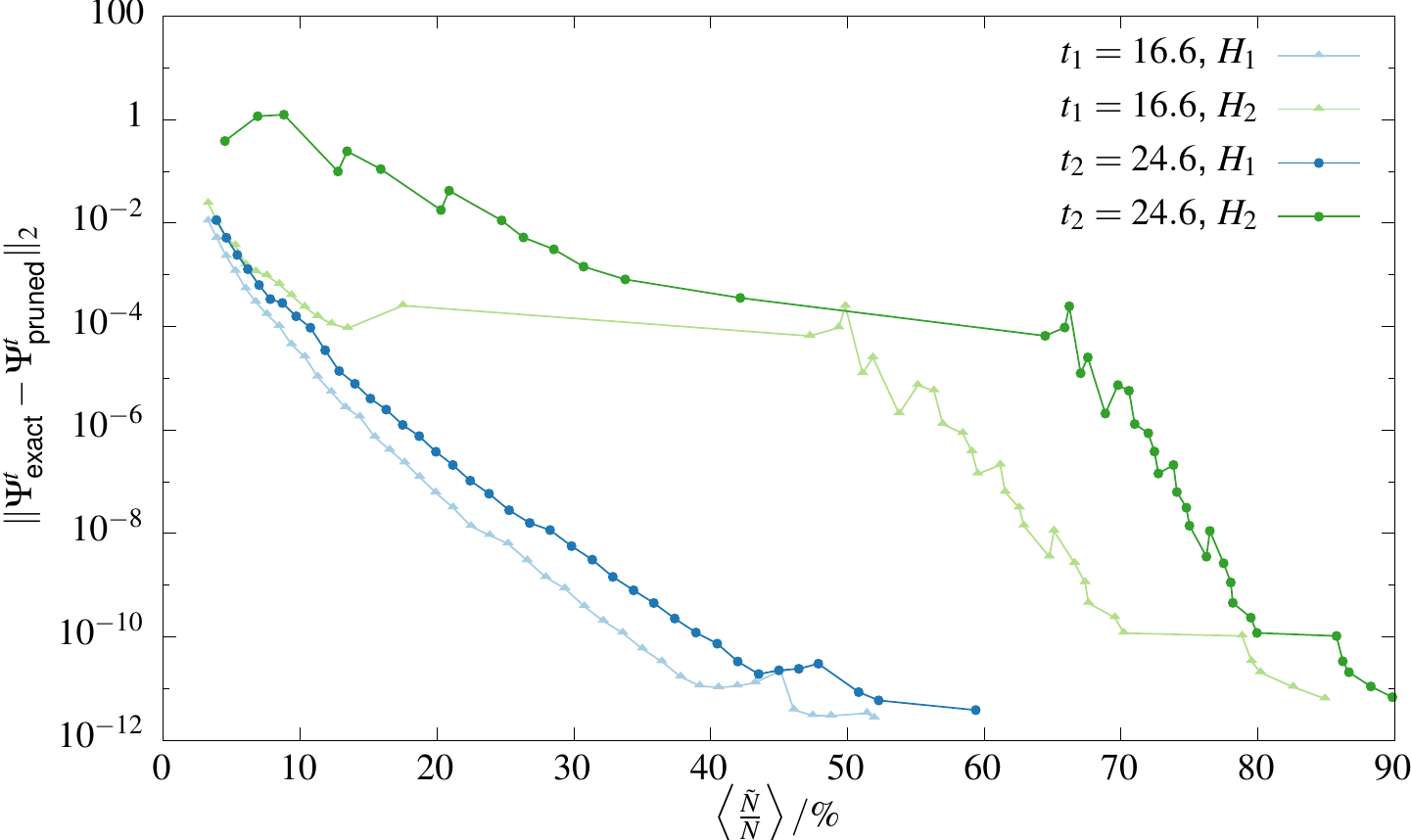}
\caption{Accuracy of the wavepacket dynamics for the 2D double-well as a function of the ratio of the reduced and unreduced basis sizes. The fidelity is shown for the two Hamiltonians, $H_1$ and $H_2$, at two times $t_1$ and $t_2$. }
\label{fig:ex_inv_comp_dwellDyn}
\end{figure}

Figure~\ref{fig:ex_inv_comp_dwellDyn} compares the accuracy of the dynamics computed using $H_1$ and $H_2$. Here we observe a significant deterioriation in $H_2$ performance. Specifically, $H_2$ fails to achieve $10^{-4}$ accuracy, both at $t_1$ and $t_2$, unless more than half of the unreduced Hilbert space is used. This is in stark contrast to $H_1$, which can achieve better than $10^{-8}$ accuracy with less than $30\%$ of the Hilbert space. Moreover, the $\approx 10^{-4}$ accuracy of $H_2$ for a very small basis size at $t_1$ is lost for $t_2$. Both aspects of this reduction in $H_2$ accuracy can be explained by recognizing the importance of the low-amplitude boundary dynamics as the wavepacket tunnels across the barrier. The Gaussians at the boundary are the first to be removed in the pruning process, as depicted in fig.~\ref{fig:deformed_Gaussians}. Therefore, $H_2$ may not be used where the low-amplitude boundary is essential for getting acceptable accuracy. In contrast, $H_1$ can be used in such situations with no difficulty, as long as the amplitude cutoff is sufficiently low so as to retain the tunneling amplitude.

\section{\label{sec:Conclusions-and-outlook}Conclusions and outlook}

In this paper we presented the mathematical underpinnings of the PvB (biorthogonal von Neumann) method for quantum mechanical simulations. The PvB method exploits the phase phase space localization of the von Neumann basis to provide a sparse representation of quantum mechanical states that spans only the part of phase space where there is significant amplitude. This in turn can lead to significant computational savings in both CPU and memory.

A detailed analysis was given of the subtle issues of projection onto subspaces of biorthogonal bases. Two complementary ways of understanding this projection were provided. The first focuses on the basis functions: it was shown that under projection one of the biorthogonal bases remains unchanged while the other becomes significantly distorted. We showed that the distortions may be viewed as arising from subtraction of the $\bar{G}$ Gaussians that span the complementary subspace. This explains why Gaussians near the boundary of the reduced phase space boundary are significantly distorted, while Gaussians far from the boundary are essentially unperturbed. The second way to understand the effect of non-orthogonal projection focuses on the coefficients: in the $\bar{G}$ basis all $\bar{B}$ basis vectors contribute to the coefficients, since all basis vectors overlap one another due to the non-orthogonality of the basis.

We then analyzed the various representations of the Schr{\"o}dinger equation in the reduced basis and approximations thereto. We concluded that for high-accuracy applications $H_1 = \widetilde{G}^\dagger H \widetilde{B}$ (eq. \ref{eq:H1_def}) is the preferred form, although it comes which a relatively high computational cost. For medium to low accuracy applications, an approximate form, $H_2 = \check{G}^\dagger H \widetilde{B}$ (eq. \ref{eq:H2}) may be used.

Several numerical examples were brought, showing the relative merits of $H_1$ and $H_2$. A more challenging application of PvB, the double ionization of helium, is presented in \cite{PvB-MCTDH-comparison-paper}.

Despite the significant methodological progress further development is possible. Specifically, one may further reduce the representation by decomposing multi-dimensional objects into a sum-of-products, and truncating the sum when the correlation is sufficiently low. This strategy is used by the POTFIT algorithm  to decompose the potential, but a similar approach could be used for the wavefunction and the reduced Hamiltonian. This is a challenging problem, however, as the dynamics continuously modify the reduced
basis, which generally is not easily decomposable.

Further areas of research include the correspondence between PvB and other phase space representations, including the discrete Husimi and Wigner representation. We also plan to explore the treatment of particle symmetries (Bosonic, Fermionic) in multi-particle implementations of the PvB method.

Beyond method development, there is a wide range of problems which are amenable to the PvB methodology, including high-harmonic generation, multi-electron ionization, photodissociation and chemical reactions. We intend to explore these applications in the near future.

To conclude, PvB is an accurate, scalable and efficient method for quantum dynamics simulations, and it is our hope that it will
find its place as part of the standard quantum numerics toolbox.

\section*{Acknowledgement}

The authors thank Tucker Carrington for useful correspondence. This work was supported by the Israel Science Foundation (533/12), the Minerva Foundation with funding from the Federal German Ministry for Education and Research and the Koshland Center for Basic Research.
H.~R.~Larsson acknowledges support by the Deutscher Akademischer Austauschdienst, Studienstiftung des deutschen Volkes and the Fonds der Chemischen Industrie.


\section*{Bibliography}

\providecommand{\latin}[1]{#1}
\providecommand*\mcitethebibliography{\thebibliography}
\csname @ifundefined\endcsname{endmcitethebibliography}
  {\let\endmcitethebibliography\endthebibliography}{}


\begin{mcitethebibliography}{34}
\bibitem[Not()]{Note-2}
Note the different definition of the norm from that in Ref.
  \cite{David-Textbook}.\relax
\mciteBstWouldAddEndPunctfalse
\mciteSetBstMidEndSepPunct{\mcitedefaultmidpunct}
{}{\mcitedefaultseppunct}\relax
\EndOfBibitem
\bibitem[Not()]{Note-3}
The transformation from the Fourier to the PvN basis can be accomplished using
  approximately $N \ln N$ operations using a Fast Gabor Transform
  \cite{Fast-Gabor-transform}, which is closely related to the Fast Fourier
  Transform.\relax
\mciteBstWouldAddEndPunctfalse
\mciteSetBstMidEndSepPunct{\mcitedefaultmidpunct}
{}{\mcitedefaultseppunct}\relax
\EndOfBibitem
\end{mcitethebibliography}
\end{document}